\documentclass[journal]{IEEEtran}

\usepackage{xcolor,soul,framed} 

\colorlet{shadecolor}{yellow}
\usepackage[pdftex]{graphicx}
\graphicspath{{../pdf/}{../jpeg/}}
\DeclareGraphicsExtensions{.pdf,.jpeg,.png}

\usepackage[cmex10]{amsmath}
\usepackage{array}
\usepackage{subfig}
\usepackage{mdwmath}
\usepackage{mdwtab}
\usepackage{eqparbox}
\usepackage{url}
\usepackage{amsfonts}
\usepackage{booktabs}
\usepackage{multirow}
\usepackage{makecell}
\usepackage{bbm}
\usepackage{stfloats}
\usepackage{caption}
\usepackage{subcaption}
\usepackage{threeparttable}
\usepackage{cite}
\usepackage{verbatim}
\usepackage{textcomp}
\usepackage{hyperref}
\hypersetup{hidelinks,
colorlinks=true,
allcolors=blue,
pdfstartview=Fit,
breaklinks=true}

\bstctlcite{IEEE:BSTcontrol}
\usepackage{graphicx}
\usepackage{etoolbox}

\begin{document}

\bstctlcite{IEEEexample:BSTcontrol}
    \title{Audio-Language Models for Audio-Centric Tasks: A Systematic Survey}
   \author{Yi Su, 
         Jisheng Bai,
       Qisheng Xu,
       Kele Xu,
       Yong Dou
 \thanks{
 This work is supported by National Science and Technology Major Project (2023ZD0121101), National University of Defense Technology (ZZCX-ZZGC-01-04).

 Yi Su, Qisheng Xu, Kele Xu and Yong Dou, are with the College of Computer Science and Technology, National University of Defense Technology, Changsha, China. Corresponding author: Kele Xu.

 Jisheng Bai is with the School of Communications and Information Engineering, Xi'an University of Posts and Telecommunications, Xi'an, China.
 }}

\markboth{IEEE Transactions on Neural Networks and Learning Systems}{IEEE Transactions on Neural Networks and Learning Systems}

\maketitle


\begin{abstract}
Audio-Language Models (ALMs), trained on paired audio-text data, are designed to process, understand, and reason about audio-centric multimodal content. Unlike traditional supervised approaches that use predefined labels, ALMs leverage natural language supervision to better handle complex real-world audio scenes with multiple overlapping events. While demonstrating impressive zero-shot and task generalization capabilities, there is still a notable lack of systematic surveys that comprehensively organize and analyze developments. In this paper, we present the first systematic review of ALMs with three main contributions: (1) comprehensive coverage of ALM works across speech, music, and sound from a general audio perspective; (2) a unified taxonomy of ALM foundations, including model architectures and training objectives; (3) establishment of a research landscape capturing mutual promotion and constraints among different research aspects, aiding in summarizing evaluations, limitations, concerns and promising directions.
Our review contributes to helping researchers understand the development of existing technologies and future trends, while also providing valuable references for implementation in practical applications.
\end{abstract}


%
\IEEEpeerreviewmaketitle




\section{Introduction}
Enabling machines to hear like humans and process audio-centric tasks has long been a significant challenge \cite{deshmukh2023pengi}. Audio-Language Models (ALMs), which are trained on audio-text data, focus on the processing, understanding, and reasoning of audio. This area is emerging as a prominent research field at the intersection of audio processing and natural language processing. ALMs are not only applicable to basic audio tasks, such as audio classification \cite{elizalde2023clap} and sound event detection, but also show great potential for more complicated scenarios. These include tasks such as audio generation \cite{liu2023audioldm}, automatic audio captioning \cite{kim2024enclap}, audio source separation \cite{liu2022_lass}, and audio chatbots \cite{zhang2023speechgpt}.

In contrast to audio representation learning based on labeled data for specific tasks, ALM learns from more descriptive textual information, expanding supervision to include human-annotated captions and web sources like titles and descriptions \cite{laion_ai_audio_dataset}. Natural language effectively characterizes real-world audio, which often involves multiple overlapping sound events, enabling models to learn their intrinsic relationships \cite{wu2019audio}. Moreover, using natural language as supervision eliminates the model's dependency on predefined task-specific labels, enhancing its potential to generalize in open-world scenarios.

Leveraging the strong comprehension capabilities of large language models (LLMs), researchers have integrated them as reasoning components into ALMs, sparking the emergence of Large Audio-Language Models (LALMs). However, pretrained LLMs still face generalization challenges across diverse downstream tasks \cite{zhao2023LLMsurvey}, requiring additional transfer steps such as post-training and collaboration with other foundation models. Within this research landscape, language offers a unified mechanism for constructing instances, enabling LLMs to undergo instruction tuning and in-context learning across various tasks. This bridges auditory information with language understanding while aligning multiple ALM components. Furthermore, language serves as a versatile human-machine interface, allowing users to direct LLM agents in coordinating effectively with audio-language systems.

Despite considerable interest in audio-language models (ALMs), no comprehensive survey currently captures the full scope of this rapidly evolving field. Existing relevant reviews include speech-language models \cite{cui2024speechlanguage, ji2024wavchat, arora2025landscape, Peng2025SLLMsurvey}, codec-based models \cite{wang2023codec}, ALMs for specific tasks such as audio-text retrieval \cite{koepke2022benchmarks}, automated audio captioning \cite{xu2023aac_survey}, speech-to-text translation \cite{xu2023S2TT}, audio-language datasets \cite{wijngaard2024ald_survey}, udio reasoning tasks \cite{guo2026reasoning}, trustworthiness of LALMs \cite{luo2026trustworth}, and evaluations of LALMs \cite{yang2025evaluationsurvey}. While these reviews offer valuable insights into ALM subdomains, their narrow focus fails to capture broader developments and cross-domain synergies. This fragmentation motivates our work: the first comprehensive ALM survey. Our three key contributions include: (1) adopting a general audio-centric perspective to cover diverse works on human voices, natural sounds, and music, offering a complete picture of computer audition progress; (2) providing a unified taxonomy of ALM foundations through systematic review, including architectures and training objectives; (3) establishing a training-oriented research landscape that captures mutual promotion and constraints among different research aspects (from model to data), aiding in systematic summarization of evluations, limitations, concerns and future directions.

\begin{figure*}
  \centering
  \includegraphics[width=0.95\textwidth]{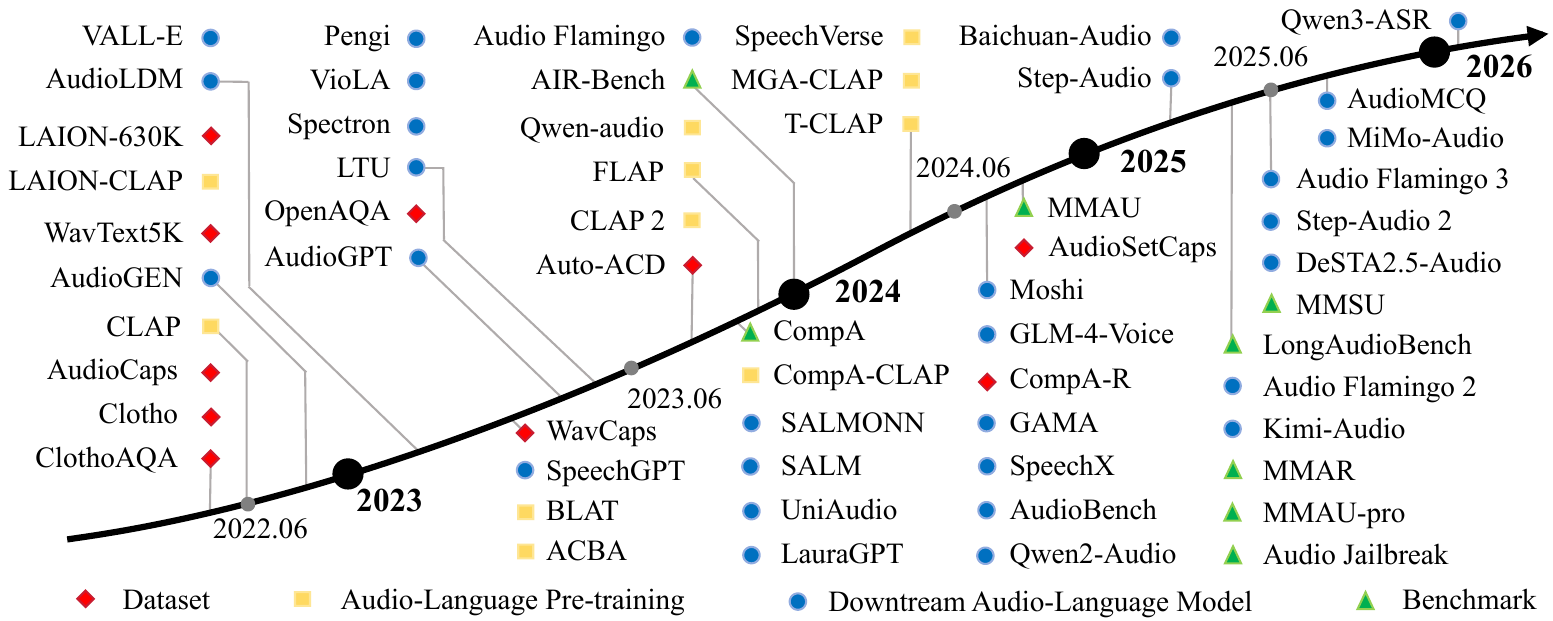}
  \caption{A timeline of recent advances in audio-language models, organized primarily by release dates (e.g., arXiv submission) with some works still in progress. It highlights that datasets serve as the foundation for pre-training and downstream research. As model research evolves, recent studies have developed multiple benchmarks to drive comprehensive field development.}
  \label{fig1: timeline}
\end{figure*}

To contextualize our contributions, Fig. \ref{fig1: timeline} illustrates the evolution of ALMs, highlighting key developments across research aspects. Early audio-caption datasets \cite{kim2019audiocaps, drossos2020clotho, lipping2022clotho} enabled significant pre-training advances like CLAP \cite{elizalde2023clap}. Subsequent large-scale datasets \cite{wu2023large} demonstrated ALMs' generalization capacity, accelerating progress in downstream tasks and benchmarks. This progression aligns with our research landscape (Fig. \ref{fig2: framework}), where pre-training, transfer learning, datasets, and benchmarks interact synergistically. We also note growing emphasis on speech applications, though general audio modeling remains challenging due to diverse sound events \cite{chen2023beats}.

In the following sections, we first present the background of audio-language pre-training and transfer learning (Section \ref{section: Background}), followed by foundations of ALMs (\ref{section: Foundations}). We then examine representation pre-training (\ref{section: Pre-training}), downstream transfer (\ref{section: transfer}), along with related datasets (\ref{section: Data}) and evaluations (\ref{section: Evaluations}). Based on this review, we discuss key limitations, concerns, and future directions (\ref{section: Limitations and Concerns}, \ref{section: Future directions}), and conclude the paper (\ref{section: Conclusions}).

\section{Background} \label{section: Background}


\subsection{Pre-training and Transfer Paradigm}
The pre-training and transfer paradigm involves training on large-scale public datasets to learn robust representations, which are then adapted to downstream tasks. While this approach accelerates supervised learning, it faces two key challenges. First, models may overfit by relying on superficial label correlations rather than learning meaningful audio content, limiting generalization \cite{gong2024ltu}. Second, the high cost of manual annotation constrains the scale of labeled data, hindering representation learning \cite{sun2024AAA}.
To address these, ALMs leverage natural language as supervision. This offers richer audio descriptions, such as temporal relationships between events (e.g., ``simultaneous" or ``before"), enabling models to better understand complex auditory scenes \cite{ghosh2024compa, wu2019audio}. Additionally, audio-text data is more scalable. While a dog barking might be variably described as "dog" or "barking," leading to annotation inconsistencies, ALMs are able to extract consistent semantic features from diverse descriptions. Beyond human-annotated texts, web-sourced titles and descriptions further expand available annotations.

\subsection{Audio-Language Training Stages}
As data volumes and model complexity grow, ALM training strategies have become increasingly sophisticated. We distinguish two main stages from representation learning and downstream application: pre-training develops task-agnostic audio representations, while transfer learning adapts models through fine-tuning and cooperative methods.
Pre-training typically involves multiple phases: initial foundation model training, audio-language alignment on paired data, and potentially extended training with diverse datasets and tasks.
Despite strong zero-shot retrieval capabilities, transfer learning remains essential for most downstream applications. Task-specific fine-tuning represents the most common approach, often incorporating adaptive modules. Multi-task transfer leverages shared knowledge across tasks through joint training. Beyond task-specific optimization, instruction tuning and in-context learning enhance LLMs’ instruction-following abilities through natural language prompts \cite{wei2022instruction}, thereby improving generalization. Multi-task transfer can also be achieved via collaborative agent systems.

\subsection{Research Landscape}
\begin{figure}[htbp]
  \includegraphics[width=0.48\textwidth]{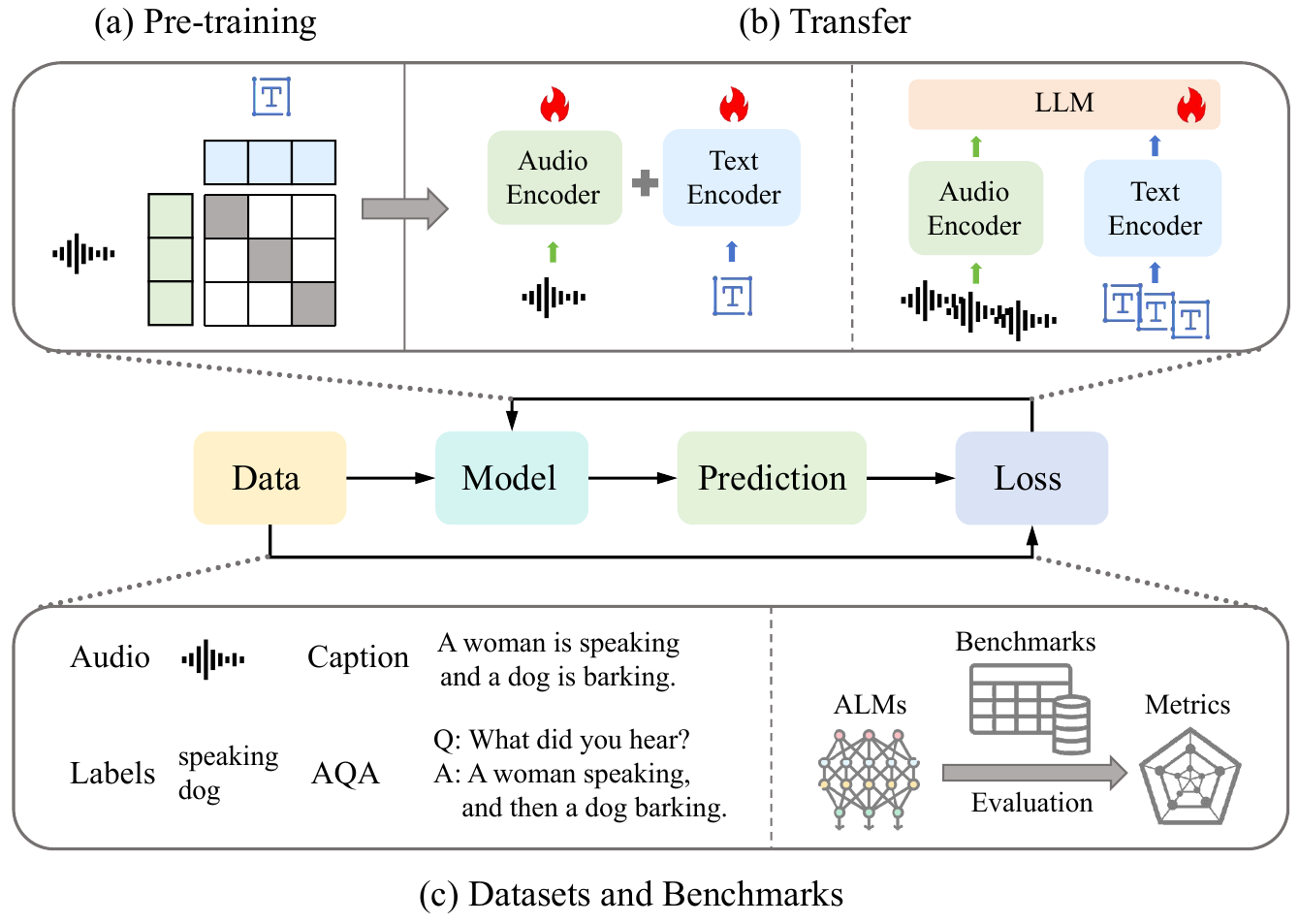}
  \caption{Research Landscape: (a) model pre-training for audio-language representations (Sec. \ref{section: Pre-training}), (b) model transfer via task-specific fine-tuning or multi-task cooperation (Sec. \ref{section: transfer}), (c) foundational data including audio-text pairs, QA datasets, and evaluation benchmarks (Sec. \ref{section: Data}).}
  \label{fig2: framework}
\end{figure}

To systematically survey ALMs, we introduce the research landscape in Fig. \ref{fig2: framework}, which captures interconnections across research areas through two core stages: (a) Pre-training, which integrates pre-trained audio and language models and aligns them using large-scale audio-text data to achieve multimodal perception capabilities; and (b) Transfer, which adapts and combines pre-trained models with other components to support diverse downstream applications. The development of both stages relies on (c) Data and Benchmarks, where datasets provide resources for model training, while benchmarks establish unified standards for evaluation and optimization.

\begin{figure*}
    \centering
    \includegraphics[width=1\linewidth]{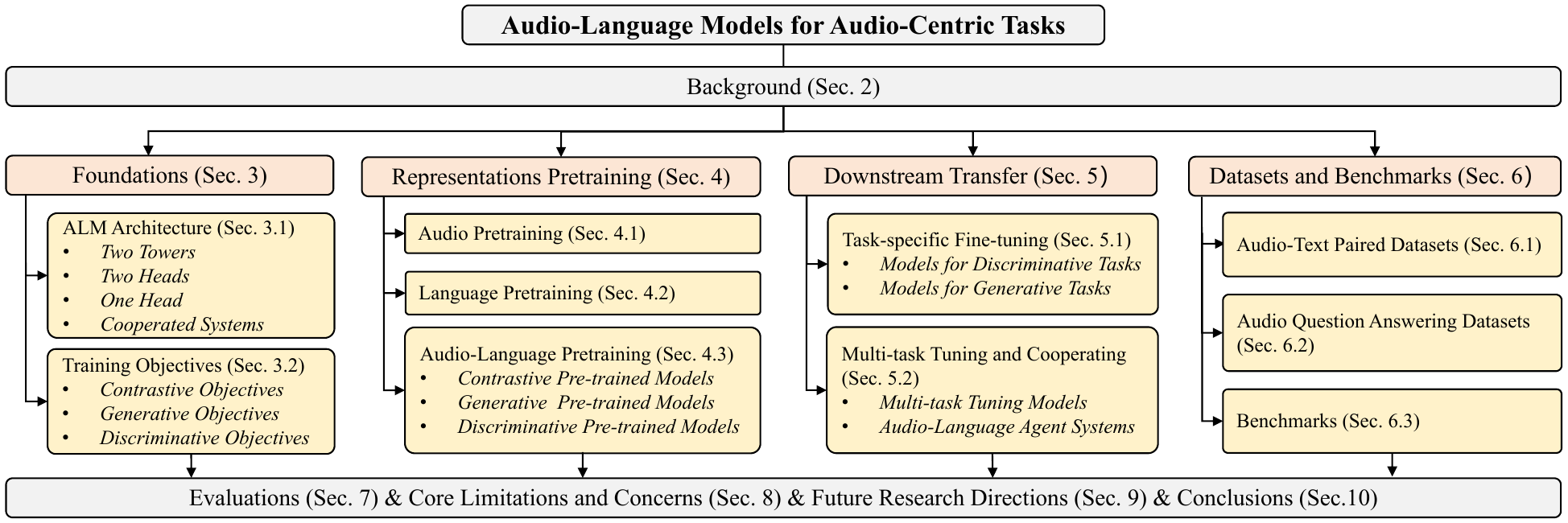}
    \caption{Research outline on audio-language models for audio-centric tasks.}
    \label{fig3:outline}
\end{figure*}

Guided by this landscape, our review outline in Fig. \ref{fig3:outline} follows a systematic progression: analyzing ALM foundations, reviewing advances across core stages, examining how data and benchmarks drive progress, and culminating in a synthesis of cross-domain challenges and future directions.

\section{Foundations} \label{section: Foundations}

\subsection{ALM Architectures}
\begin{figure*}
    \centering
    \includegraphics[width=0.93\linewidth]{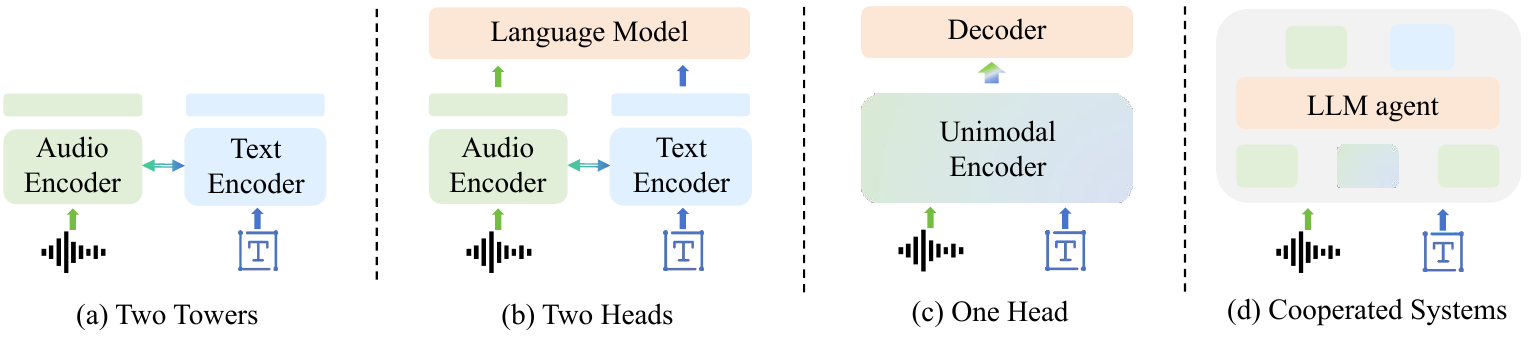}
    \caption{Typical architectures of audio-language models. (a) Two Towers, with one encoder and projector for each modality, embeddings will be aligned in a joint space. (b) Two Heads, adds language model on top. (c) One Head, with a unimoal encoder and decoder. (d) Cooperated Systems, utilize LLMs as agents to cooperate several models.}
    \label{fig4:arch}
\end{figure*}

ALMs typically comprise audio and language encoders, and may include multimodal alignment mechanisms and language modeling networks. As shown in Fig. \ref{fig4:arch}, we categorize ALMs into four main types: Two Towers, Two Heads, One Head, and Cooperated Systems. These architectures, defined by distinct data flow patterns embody different multimodal philosophies, and directly influence deployment and performance.

\subsubsection{Two Towers} 
This basic architecture employs separate encoders and projectors for each modality, aligning embeddings in a joint space. Exemplified by the landmark CLAP model, it uses contrastive learning to map audio and text into a shared multimodal representation \cite{elizalde2023clap}. This late-interaction design yields low inference complexity, with generalization strengths in cross-modal retrieval and zero-shot classification. By shifting online inference to similarity computation through pre-computed cached representations, Two-Tower architectures achieve low-latency deployment for large-scale retrieval. While typically requiring hundreds of millions of parameters, memory can be reduced via knowledge distillation \cite{paissan2024tinyclap}. Alignment mechanisms can be integrated between encoders to enhance representation learning \cite{li2021ctal}.

\subsubsection{Two Heads} 
This mainstream architecture adds a language model atop separate modality encoders and projectors. The 'Head' projects modal representations into a unified space \cite{jang2023oner}. Language modeling enhances speech tasks \cite{wang2023VALLE}, while evolving LLMs serve as reasoning backbones \cite{deshmukh2023pengi, chu2023qwenaudio}. Modality fusion can be enhanced through encoder alignment mechanisms \cite{zhao2024mint}, and text may be directly tokenized without a dedicated encoder \cite{das2024speechverse}. This intermediate-fusion architecture, inference latency and memory requirements are dominated by the LM, varying with model scale and generation mode (e.g. autoregressive/non-autoregressive). Internal variation across different LM choices is substantial. Initialization with powerful LMs aids convergence and generalization via knowledge transfer \cite{kaplan2020scaling}, though stability depends on module selection and fine-tuning strategies.

\subsubsection{One Head}
This unified architecture processes both modalities through a single encoder before decoding. While vision research shows shared multimodal processing improves alignment \cite{jang2023oner}, audio-language applications remain limited \cite{sachidananda2022calm}. This early-fusion approach has primarily explored using joint representations for downstream task fine-tuning \cite{sachidananda2022calm}, with no existing work incorporating LLMs for decoding. Inference latency in such models primarily comes from the fusion module's forward computation. Theoretically, a unimodal encoder has the potential to achieve faster inference efficiency than Two Heads architectures, but faces steeper convergence challenges from joint cross-modal optimization in a unified parameter space \cite{xia2023unicode}.

\subsubsection{Cooperated Systems}
This system employs an LLM as a planning agent and comprises various model types mentioned above. It selects and facilitates different models based on their complementary strengths. Through the collaboration of these diverse models, the system can tackle a wider array of complex tasks compared to a solitary model alone \cite{li2024survey}. Their memory usage varies with the loaded foundation models, and inference latency depends on execution paths. They face unique convergence challenges from the interplay between component performance and policy learning, and achieve distinctive generalization through modular capability composition.

Given the distinct characteristics of each architecture, the choice of which to adopt should be guided by application requirements. Two-tower models excel in retrieval-heavy scenarios that benefit from precomputed embeddings; two-head models are preferable for open-ended reasoning and instruction following; one-head models remain promising for unified modeling but are less mature; cooperated systems offer flexibility for tasks requiring modular decomposition or external tools. These choices must also consider the constraints of different deployment environments: cloud-based batch processing can accommodate large multimodal models for high accuracy; edge or on-device inference, being resource-constrained, is better suited for two-tower models or lightweight codec-based models; real-time streaming systems, sensitive to latency, require careful management of decoding overhead and bitrate-fidelity trade-offs.

\subsection{Training Objectives}
\begin{figure*}[ht]
    \centering
    \includegraphics[width=0.97\textwidth]{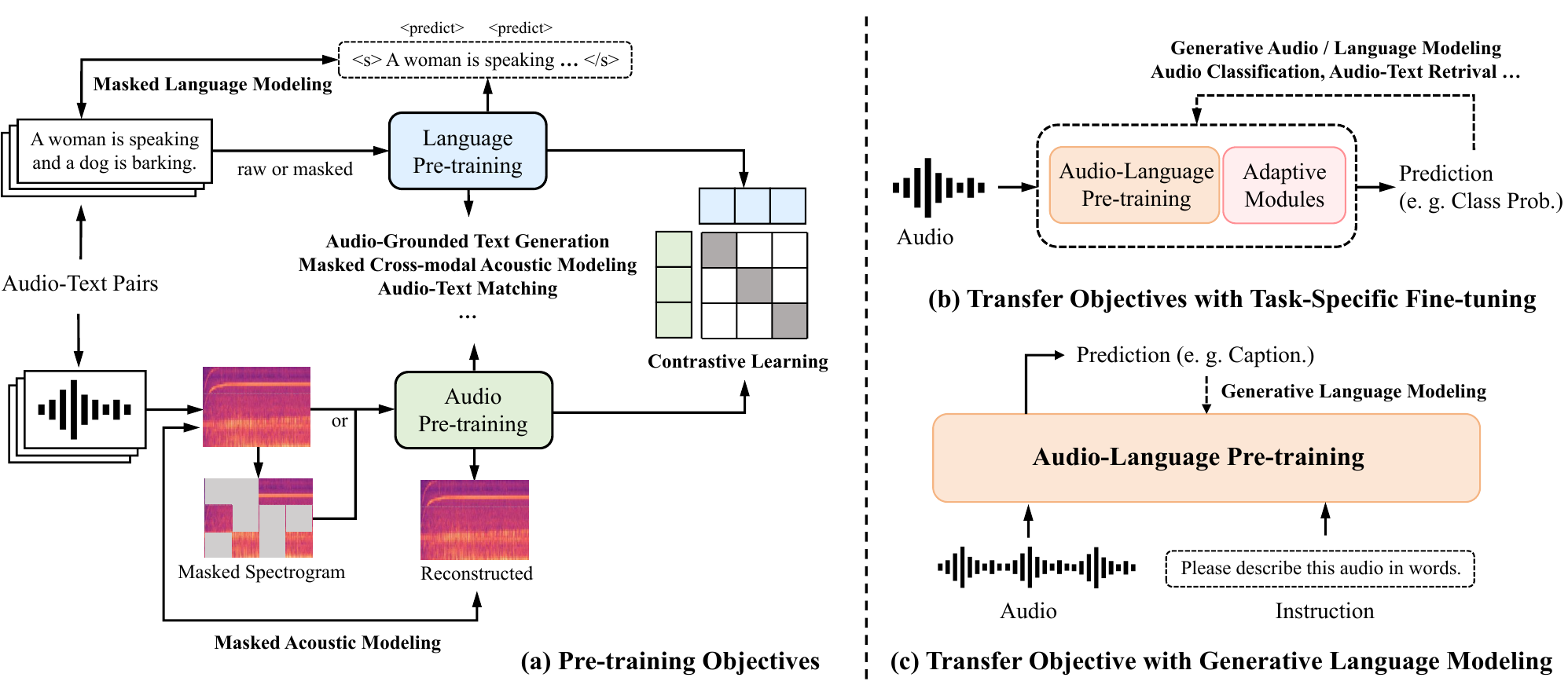}
    \caption{Illustration of audio-language models training objectives. (a) Pre-training objectives include contrastive, generative, and discriminative objectives, which may be conducted on audio-text or single-modal data. The transfer objectives can be (b) task-specific fine-tuning objectives or (c) generative language modeling objective.}
    \label{fig5: objectives}
\end{figure*}

Training objectives guide model learning during pre-training and transfer. As shown in Fig. \ref{fig5: objectives}(a), pre-training objectives include contrastive, generative, and discriminative losses that learn pretext tasks on audio, text, or audio-text pairs to capture semantic features and cross-modal correlations. Fig. \ref{fig5: objectives}(b) illustrates task-specific fine-tuning using generative or discriminative objectives, while Fig. \ref{fig5: objectives}(c) shows transfer via generative language modeling. Note that the choice of training objective depends on the specific requirements of the learning stage and application scenario, and the above training objectives can be used in combination. Theoretically, contrastive objectives are well-suited for improving cross-modal alignment, and matching loss are generally designed for finer-grained alignment. Generative objectives, including reconstruction and language modeling losses, are advantageous when fine-grained acoustic grounding or open-ended content generation is required, and reconstruction losses can stabilize training under noisy or limited textual supervision.
For detailed ablation studies on objective combinations, see works like MINT \cite{zhao2024mint, yeh2023flap}.

\subsubsection{Contrastive Objectives}
Contrastive learning, widely used in audio-language pre-training, brings positive audio-text pairs closer while pushing negatives apart in a shared embedding space, thereby learning the audio-language correlations and obtaining distinguishable representations between audio samples \cite{elizalde2023clap}. The most widely implemented approach for this category of objective is using a symmetric audio-text infoNCE \cite{oord2018CPC} loss function to measure the similarity between audio and text embeddings.  
For a batch of $B$ pairs $\{x_i, t_i\}$, with audio encoder $h_a(\cdot)$ and text encoder $h_t(\cdot)$, embeddings are:
\begin{equation}
z_{i}^{a}=h_a(x_i), \quad z_{i}^{t}=h_t(t_i)
\end{equation}
The similarity between audio and text embeddings is computed via dot product. For a batch of $B$ audio-text pairs, the loss for the audio dimension $l_{i}^{a}$ measures how well the audio query $z_i^a$ matches its corresponding text $z_i^t$ against all other texts in the batch. Similarly, the text dimension loss $l_{i}^{t}$ assesses the text query $z_i^t$ against all other audios:
\begin{equation}
l_{i}^{a}=-\log \frac{\exp \left(z_{i}^{a} \cdot z_{i}^{t} / \tau\right)}{\sum_{j=1}^{B} \exp \left(z_{i}^{a} \cdot z_{j}^{t} / \tau\right)}
\end{equation}
\begin{equation}
l_{i}^{t}=-\log \frac{\exp \left(z_{i}^{t} \cdot z_{i}^{a} / \tau\right)}{\sum_{j=1}^{B} \exp \left(z_{i}^{t} \cdot z_{j}^{a} / \tau\right)}
\end{equation}
Here, $\tau$ is a temperature parameter scaling the logits. The symmetric contrastive loss combines both directions:
\begin{equation}
\mathcal{L}_{con}=\frac{1}{2B} \sum_{i=1}^{B}(l_{i}^{a} + l_{i}^{t}) 
\label{eq:infonce}
\end{equation}

Theoretically, contrastive loss is non-decomposable, since the loss for a positive pair couples with all negatives in the batch. This implies that a finite batch size limits the negative pool to a sparse, biased subset of the true distribution, introducing systematic gradient bias and suboptimal convergence \cite{chen2022gradientbias}. While ALMs commonly adopt large batches (e.g., 2304 in FLAP \cite{yeh2023flap}) and have explored hard negative mining \cite{li2024MGACLAP} to mitigate sampling bias, the problem itself remains underexamined in the community. Large-batch training improves performance but incurs high computational costs. More principled solutions from contrastive learning, such as momentum encoders \cite{he2020moco} and debiased techniques based on importance weighting \cite{chuang2020debiased}, decouple negative pool size from batch constraints or correct sampling bias through weighting, enabling more accurate estimation of the true contrastive loss. We highlight these as promising yet unexplored directions for future contrastive ALM research.

\subsubsection{Generative Objectives}
Generative methods effectively learn audio representations through pretext tasks like masked reconstruction of audio and associated text \cite{huang2022audiomae}. These methods are often combined with contrastive learning, they enhance embedding robustness and computational efficiency. In transfer learning, these objectives adapt models to downstream generation tasks and are widely used in generative LLM scenarios.


During pre-training, the most common method for audio mask reconstruction operates on spectrograms. Let $M(\cdot)$ denote the masking operation, $f_a(\cdot)$ the spectrogram encoder, and $p_{ae}(\cdot)$ the audio embedding projection layer. Masked spectrogram prediction incorporates an additional decoder $f_a^{-1}(\cdot)$. For an original spectrogram $a$, the reconstruction is formulated as $\hat{a} = f_a^{-1}(p_{ae}(f_a(M(a))))$. Denoting the decoder's prediction and the original true value for the $n$-th masked spectrogram patch as $\hat{a}_n$ and $a_n$ respectively, the self-supervised audio reconstruction loss for $N$ patches minimizes the $L2$ (mean squared error, MSE) loss:
\begin{equation}
\mathcal{L}_{ar}=\frac{1}{N}\sum _{n=1}^{N}\left\|\hat{a}_{i}-a_{i}\right\|_{2}
\label{eq:rec}
\end{equation}
As ALMs incorporate both audio and language modalities, some works design masked cross-modal reconstruction tasks. These typically use cross-attention mechanisms to enable communication between the two encoders and reconstruct the audio representation.

When conducting transfer through audio generation tasks, the training minimizing the distance between the predicted embedding $\hat{z}$ and ground truth $z$. Common distance metrics include $L1$ and $L2$ norms, with the objective often formulated as their weighted sum:
\begin{equation}
\mathcal{L}_{am}=\frac{1}{T} \frac{1}{L} \sum_{t=1}^{T} \sum_{l=1}^{L} \alpha \left\| \hat{z}_{t, l}-{z}_{t, l}\right\|_{1}+\beta \left\| \hat{z}_{t, l}-{z}_{t, l}\right\|_{2}
\label{eq:am1}
\end{equation}
where $T$ is the total frames, $L$ the embedding dimension, and $\alpha$, $\beta$ are weight hyperparameters.
Alternatively, the objective can be applied directly to audio amplitude. For a decoder network $h_{de}(\cdot)$ that maps input audio $x_i$ and text query $t_i$ to predicted audio $|\hat{a}_i|$,  with $z_i^t$ as the language embedding, the $L1$ loss between amplitude spectrograms is:
\begin{equation}
|\hat{a}_i| = h_{de}\left (z_{i}^{t} \right ), \quad \mathcal{L}^{'}_{am} = \sum_{i=1}^B \left \| |{a}_i| - |\hat{a}_i| \right \|_{1}
\label{eq:am2}
\end{equation}

Generative language modeling objectives guide ALMs to produce audio-related text consistent with ground truth. These objectives help learn audio-language correlations for improved performance on downstream tasks (e.g., automatic captioning) and it serve as standard losses during ALM transfer with language models \cite{JMLR2023palm}.

Language generation requires additional text decoder (e.g., a pre-trained language model), where the autoregressive objective minimizes the negative log-likelihood of each ground-truth token given the preceding tokens and audio input $x$:
\begin{equation}
\mathcal{L}_{lm}=-\frac{1}{T} \sum_{t=1}^{T} \log P_{\theta}\left(y_{t} \mid y_{1: t-1}, x\right)
\label{eq:lm_obj}
\end{equation}
Here $y_t$ is the $t$-th ground-truth token of caption $y$, $T$ is the caption length, and $\theta$ represents model parameters. Non-autoregressive models use a similar negative log-likelihood objective without temporal averaging.

\subsubsection{Discriminative Objectives}
They are used to guide the model in learning to predict the correct label, and can be broadly categorized into classification and retrieval objectives. Here, we take the cross-entropy function as an example to uniformly calculate the loss between the predicted output and the ground truth.

Audio classification is one of the most extensively studied downstream tasks. It aims to recognize patterns from specific audio inputs to predict given labels. For a batch of $B$ audio samples, the objective can be expressed as:
\begin{equation}
\mathcal{L}_{cls} = -\frac{1}{B} \sum_{i=1}^{B} \sum_{c=1}^{C} y_{i,c} \log(\hat{p}_{i,c})
\label{eq:cls}
\end{equation}
where $C$ is the number of classes. $y_{i,c}$ is the true label of the $i-th$ sample in class $c$ (0 or 1). $\hat{p}_{i,c}$ is the predicted probability of the $i-th$ sample in class $c$.

Audio-Text Retrieval (ATR) aims to find matching items between audio clips and textual descriptions. Given a query in one modality (audio or text), it retrieves the corresponding item from candidate pools in the other modality. Here, we use a scoring function $S\left (\cdot \right )$ to represent the model's prediction output by measuring the correlation between audio and text. 
Denote $Y$ as a set of $m$ possible caption texts, the correspondence caption of a given audio $x_i$ is
\begin{equation}
\hat{y}_i = \arg\max_{y_j \in Y} \frac{\exp(S(z_{i}^{a}, z_{j}^{t}))}{\sum_{k=1}^{m} \exp(S(z_{i}^{a}, z_{k}^{t}))}
\end{equation}
Then, retrieval tasks can be considered as instance-level classification, so the objective can be formatted as:
\begin{equation}
\mathcal{L}_{atr} = - \sum_{i=1}^B \log(\hat{y}_i)
\end{equation}
Specifically, audio-text matching is a pretext task designed to force a more fine-grained alignment between audio and text embeddings than contrastive pre-training. This task trains the model to predict whether a given text correctly describes a provided audio segment, functioning as a binary classification task that determines whether an audio-language pair matches or not. The matching objective can be defined as:
\begin{equation}
\mathcal{L}_{mat}=p \log \mathcal{S}\left(z^{a}, z^{t}\right)+(1-p) \log \left(1-\mathcal{S}\left(z^{a}, z^{t}\right)\right)
\end{equation}
Here, $p$ is 1 if the audio and text are paired, otherwise it is 0.

\section{Representation Pre-training} \label{section: Pre-training}
In recent years, the pursuit of powerful audio representations has led to significant growth in both audio dataset sizes and model scales. Training ALMs becomes more complex, encompassing multiple stages of pre-training aimed at enhancing task-independent audio representations before transfer to downstream tasks.

Audio encoders and text encoders are the most critical components of ALMs. They provide the initialization of model parameters for post pre-training on audio-text pairs or transfer on downstream task datasets. Recent studies have shown that the choice of encoders significantly impacts the generation of powerful representations through Audio-language pre-training and enhances the performance of downstream tasks \cite{mei2024wavcaps}.

\subsection{Audio Pre-training}

From the perspective of model architecture, pre-trained models used to initialize encoders in ALMs mainly include CNN-based, Transformer-based, and codec-based models. CNN-based models like PANNs \cite{kong2020panns} extract features from spectrograms and are widely used in audio classification. Transformer-based models dominate current research, with representative works including Wav2vec 2.0 \cite{baevski2020wav2vec}, HuBERT \cite{hsu2021hubert}, Whisper \cite{radford2023whisper}, AST \cite{gong2021ast}, and AudioMAE \cite{huang2022audiomae} for speech and general audio tasks. Codec-based models such as SoundStream \cite{zeghidour2021soundstream} and Encodec \cite{fossez2023high} discretize audio via encoder-decoder structures for language modeling. Due to space limitations, detailed introductions of these models are provided in Supplementary Material.

\subsection{Language Pre-training}

Language pre-training provides foundational text encoders for ALMs, enabling them to process and generate natural language descriptions related to audio content. These encoders are typically initialized with pre-trained language models before being jointly optimized with audio encoders during audio-language pre-training. Popular families include GPT \cite{radford2019GPT2, achiam2023gpt4}, LLaMA \cite{touvron2023llama2, chiang2023vicuna}, Qwen \cite{bai2023qwen}, and OPT \cite{zhang2022opt}. For comprehensive reviews of language pre-training, we refer readers to \cite{zhao2023LLMsurvey, minaee2024LLMs}.

\subsection{Audio-Language Pre-training}
From model training perspective, the objectives in audio-language pre-training can generally be categorized into three types: contrastive, generative objectives, and discriminative objectives. Table \ref{tab1:ALPs} shows the audio-language pre-trained models and objectives they use.

\begin{table}[htbp]
    \centering
    \caption{Summary of Audio-Language Pre-training Models.}
        \begin{tabular}{ccccc}
        \toprule
        \multirow{2}{*}{Model} & \multirow{2}{*}{Pre-trained Models} & \multicolumn{3}{c}{Objectives} \\
        &  & Con & Gen & Dis \\
        \midrule
        MS-CLAP \cite{elizalde2023clap} & CNN14+BERT & \checkmark &  &  \\
        LAION-CLAP \cite{wu2023large} & HTSAT+RoBERTa & \checkmark & &  \\
        MS-CLAP V2 \cite{elizalde2024CLAP2} & HTSAT-22+GPT-2 & \checkmark & & \\
        BLAT \cite{xu2023blat} & CNN14+BERT  & \checkmark & & \\
        ACBA \cite{wu2023ACBA} & CNN14+RoBERTa & \checkmark & & \\
        COMPA \cite{ghosh2024compa} & HTSAT+Flan-T5 & \checkmark & & \\
        MGA-CLAP \cite{li2024MGACLAP} & HTSAT/AST+BERT  & \checkmark & & \\
        T-CLAP \cite{yuan2024tclap} & HTSAT+RoBERTa  & \checkmark & & \\
        MusCALL \cite{manco2022contrastive} & ResNet50+GPT-2  & \checkmark & & \\
        \midrule
        CTAL \cite{li2021ctal} & RoBERTa &  & \checkmark &  \\
        FLAP \cite{yeh2023flap} & MAViL+RoBERTa & \checkmark & \checkmark & \\
        M2D-CLAP \cite{niizumi2024m2dclap} & M2D+GTE & \checkmark & \checkmark & \\
        Cacophony \cite{zhu2024cacophony} & AudioMAE+RoBERTa & \checkmark & \checkmark & \\
        \midrule
        MINT \cite{zhao2024mint} & Data2vec+Flan-T5 & \checkmark & \checkmark & \checkmark \\
        \bottomrule
        \end{tabular}%
    \label{tab1:ALPs}%
    \begin{tablenotes}
        \item[] Objectives: Contrastive (Con), Generative (Gen), Discriminative (Dis).
    \end{tablenotes}
\end{table}%

\subsubsection{Contrastive Pre-trained Models} Contrastive learning is a common pre-training method that learns feature representations by distinguishing similar sample pairs (positive pairs) from dissimilar ones (negative pairs). Inspired by the CLIP \cite{RadfordCLIP2021} model, which leverages large-scale web visual-text pairs to expand the training scale for visual tasks and achieves excellent performance across tasks, researchers have apply this contrastive training paradigm to the audio field. Microsoft's MS-CLAP \cite{elizalde2023clap} (prefix added by the editor for distinction) is the first contrastive language-audio pre-training model. It uses the symmetric audio-text infoNCE loss function \ref{eq:infonce} for pre-training based on audio-text paired datasets and audios from other tasks such as audio classification.

Subsequent studies have focused on dataset scaling. LAION-CLAP \cite{wu2023large} released a larger audio-caption dataset, LAION-Audio-630K, and trained the first fully open-source CLAP model together with other public datasets. MS-CLAP V2 \cite{elizalde2024CLAP2} not only leverages a more extensive multi-task trained audio encoder but also further expands the audio-text paired dataset for contrastive learning. BLAT \cite{xu2023blat}, from another prospective, proposes using an audio captioning model to generate audio-text pair data for contrastive pretraining.

Another line of research aims to address the inherent shortcomings of the vanilla CLAP. The experiments of ACBA \cite{wu2023ACBA} show that CLAP has limited understanding of natural language, especially regarding the order or concurrent arrangement of sound events. It suggests modifying the original pre-training dataset to provide more audio-language pairs about ordering. COMPA \cite{ghosh2024compa} addresses the issue that current benchmarks cannot measure the lack of combinatorial reasoning in models and proposes contrastive training with composition-aware hard negatives and a modular contrastive loss to improve combinatorial reasoning capabilities. MGA-CLAP \cite{li2024MGACLAP} tackles the problem of different granularities between modalities by adopting a shared codebook, designing a locality-aware block and a hard-negative guided loss to achieve fine-grained alignment. T-CLAP \cite{yuan2024tclap} identifies its weakness in capturing temporal information and introduces a temporal-contrastive loss, using LLM-generated ordered and misordered captions to learn the sequence of sound events.

Additionally, research has also been conducted in specialized fields like music, with works such as MusCALL \cite{manco2022contrastive} achieving excellent performance on relevant tasks.

\subsubsection{Generative Pre-trained Models} 
Generative pre-training aims to learn deeper semantic representations by setting generative audio or text as pretext tasks. Cross-modal Transformer for Audio-and-Language (CTAL) \cite{li2021ctal} is an early exploration of audio-language pretraining through masked language modeling and cross-attention based masked cross-modal acoustic modeling. Fast Language-Audio Pre-training (FLAP) \cite{yeh2023flap} conduct representation learning through the combination of masked reconstruction and contrastive learning. It generates multiple augmented views of the audio through masking for inter-modal contrast and learns to reconstruct the masked parts of the audio spectrogram. This masking reduces the amount of data that needs to be processed, thereby lowering the computational complexity and making it more efficient than contrastive learning with raw spectrograms. Additionally, by incorporating the masked reconstruction task, the model is encouraged to compress information into each sample embedding, making the audio embedding not only close to their textual counterparts but also producing more informative original inputs. M2D-CLAP \cite{niizumi2024m2dclap} addresses the issue of MAE using all patches to encode training signals, which may lead to underutilization of inductive biases. By combining Masked Modeling Duo (M2D) \cite{niizumi2023m2d} to train the audio encoder while contrastive learning further promotes input modeling, thus enhancing the effectiveness of the learned representations. 
Practice in the vision-language has shown that integrating auxiliary captioning objectives in contrastive learning can provide stronger supervision \cite{yu2022coca, li2022blip}. Cacophony \cite{zhu2024cacophony} improves CLAP by incorporating an auxiliary captioning objective, encouraging the audio encoder to capture fine-grained patterns closely matching text descriptions.

\subsubsection{Discriminative Pre-trained Models} 
Discriminative pre-training aims to set up a pretext task of audio-text matching, allowing the model to learn cross-modal alignment features. MINT \cite{zhao2024mint} is a framework that enhances audio-language pre-training through multiple objectives. Specifically, it introduces Bridge-Net as a trainable module, taking the output of the audio encoder and text as inputs to the Bridge-Net. It uses contrastive objective to align the audio and text representations by maximizing the mutual information between them, combines matching objective for fine-grained audio-text alignment, and employs generative objective to guide the audio-grounded text generation task, forcing the model to extract audio features to capture all necessary information about generating the text.

\section{Downstream Transfer} \label{section: transfer}
Downstream transfer is crucial for enhancing ALM performance and enabling efficient adaptation to new tasks. It involves applying knowledge acquired from pretext tasks to solve related problems \cite{deshmukh2023pengi}. This process adapts general-purpose models to domain-specific needs, facilitating deployment of robust systems in real-world applications like voice assistants and audio search. Transfer methods comprise task-specific fine-tuning and multi-task tuning/cooperation.

\begin{figure*}[ht]
\centering
\includegraphics[width=0.92\textwidth]{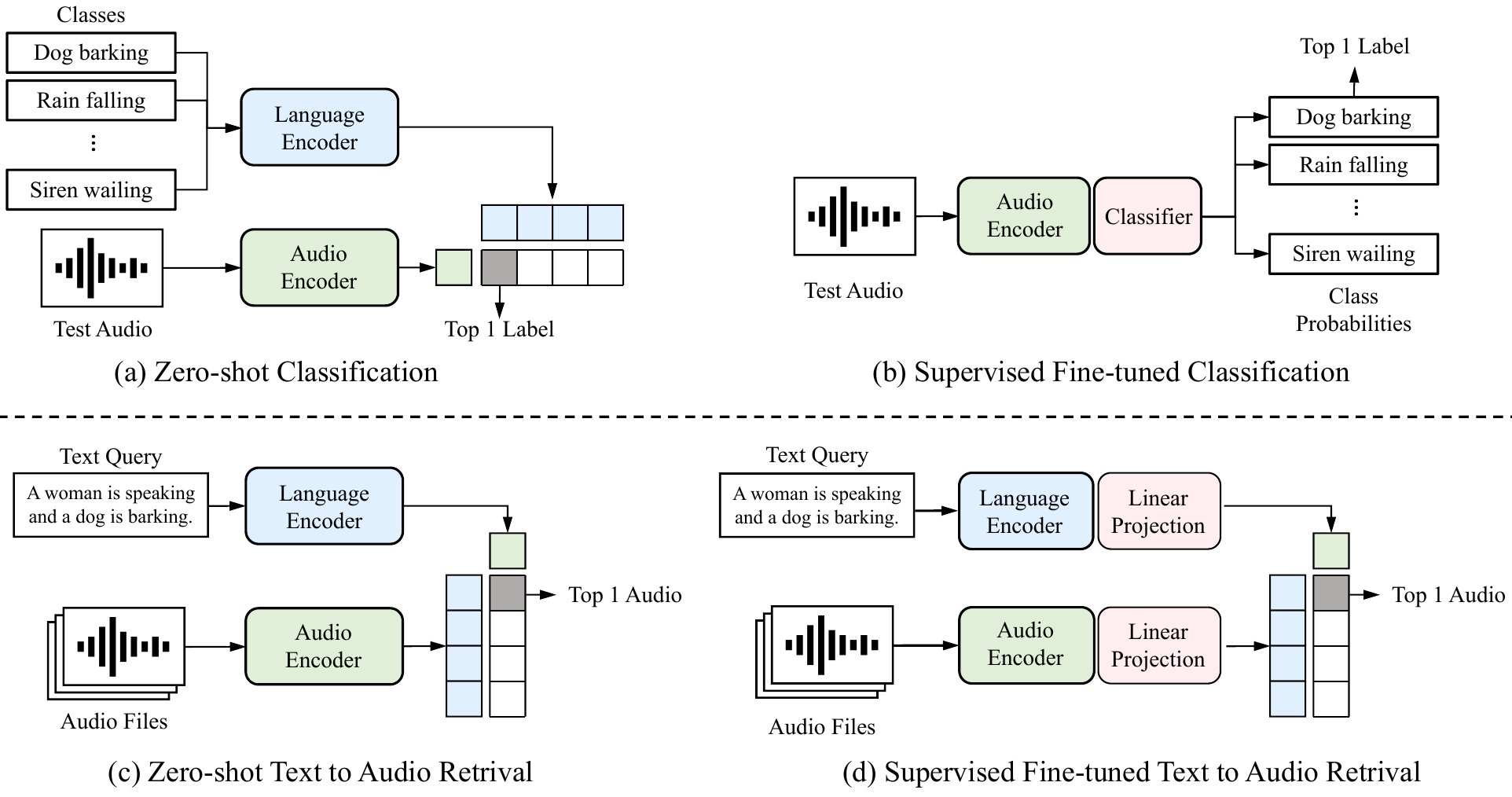}
\caption{A common framework for audio-text discriminative tasks. Pre-trained ALM can conduct classification or retrieval tasks by calculating similarities in a zero-shot way, as shown in (a) and (c). It can also connect adaptive modules and undergo supervised fine-tuning on specific datasets to be used for various discriminative tasks, such as classification as shown in (b), and retrieval tasks as shown in (d).}
\label{fig6:discriminative_models}
\end{figure*}

\subsection{Task-specific Fine-tuning}
Pre-trained audio and language models can undergo supervised fine-tuning on specific downstream task datasets. In most cases, this is an essential step to leverage pre-trained models combines with other adaptive modules, such as adding a linear projection layers to map audio representations onto a specified number of categories. Moreover, this task-specific fine-tuning bridges the gap between the source and target domains, facilitating knowledge transfer across domains.

As a result, we are seeing remarkable improvements in various audio-centric discriminative and generative audio tasks. In Table \ref{tab2:task-specific fine-tuning models}, we compare different task-specific fine-tuning models in terms of application domain, architectures or pre-trained models, downstream tasks, input and output modalities. For the application domain, we use `speech' and `music' to represent domain-specific models, and `audio' to indicate models that can handle more general audio tasks. 

Common discriminative tasks include audio classification (AC) and audio-text retrieval (ATR). Audio classification encompasses specific scenarios such as sound event detection, accent recognition, chord identification, spoken language identification, speech emotion recognition (SER), and spoken language understanding (SLU).
Generative tasks are categorized into audio generation and language generation. Typical audio generative tasks include: speech continuation (SC), text-to-speech (TTS), and speech enhancement (SE) in the speech domain \cite{wang2023VALLE, zhang2023VALLEX, xue2024mmace, anastassiou2024seedtts, chen2024f5tts, Borsos2023audiolm}; piano continuation (PC), stereo generation (SG), text-to-music (TTM) and music editing (ME) in the music domain \cite{agostinelli2023musiclm, melechovsky2024mustango, Borsos2023audiolm}; as well as text-guided audio continuation (TAC), text-to-audio (TTA), text-to-spatial-audio (TTSA, also referred as spatial audio generation), language-query sound source separation (LASS), and text-to-audio grounding (TAG, also referred as text-query sound event detection) in other general audio domains \cite{kreuk2023audiogen, liu2023audioldm, liu2022_lass, dong2023clipsep, liu2024separate, mahmud2024opensep, yuan2025flowsep, yin2024TQSED, Xu2024TAG}. Language generation, centers on transcription, translation, and caption. Typical tasks include automatic speech recognition (ASR) \cite{radford2023whisper, li2024style, bai2024seedasr}, speech-to-text translation (S2TT) \cite{zhang2023VALLEX}, speech-to-speech translation (S2ST) \cite{radford2023whisper}, and automatic audio captioning (AAC) \cite{doh2023lp, mei2021audio, kim2024enclap, Kim2024enclap++, Ghosh2024Recap, li2024drcap}. 

\begin{table*}[htbp]
  \centering
  \small
  \caption{Summary of Task-specific Fine-tuning Models.}
    \begin{tabular}{cccccccc}
    \toprule
    \multirow{2}[3]{*}{Model} & \multirow{2}[3]{*}{Domain} & \multirow{2}[3]{*}{Architectures / Pre-trained models} & \multirow{2}[3]{*}{Tasks} & \multicolumn{2}{c}{Input} & \multicolumn{2}{c}{Output} \\
\cmidrule{5-8}          &       &       &       & A     & T     & A     & T  \\
    \midrule
    VALL-E \cite{wang2023VALLE} & \multicolumn{1}{c}{\multirow{9}[2]{*}{Speech}} & EnCodec+AR/NAR LM & TTS   & \checkmark     & \checkmark     & \checkmark     &  \\
    VALL-E X \cite{zhang2023VALLEX} &       & EnCodec+AR/NAR LM & TTS, S2ST & \checkmark     & \checkmark     & \checkmark     &  \\
    Seed-TTS \cite{anastassiou2024seedtts} &       & Speech Tokenizers+AR LM+DiT+Vocoders & TTS   & \checkmark     & \checkmark     & \checkmark     &  \\
    F5-TTS \cite{chen2024f5tts} &       & ConvNeXt+DiT & TTS   & \checkmark     & \checkmark     & \checkmark     &  \\
    Whisper \cite{radford2023whisper} &       & BPE + encoder-decoder Transformer & ASR, S2TT & \checkmark     &       &       & \checkmark \\
    Seed-ASR \cite{bai2024seedasr} &       & LUISE+MoE LLM & ASR   & \checkmark     & \checkmark     &       & \checkmark \\
    Qwen3-ASR \cite{shi2026qwen3asr} &       & AuT Encoder+Qwen3-Omni & ASR   & \checkmark & \checkmark &       & \checkmark \\
    LLaSE-G1 \cite{kang2025llaseg1} &       & WavLM+LLaMA+Xcodec2 & SE   & \checkmark     &      &       & \checkmark \\
    \midrule
    MusicLM \cite{agostinelli2023musiclm} & \multirow{4}[2]{*}{Music} &  SoundStream+w2v-BERT+MuLan & TTM   & \checkmark     & \checkmark     & \checkmark     &  \\
    MusicGen \cite{copet2024simple} &       & EnCodec+AR LMs & TTM, SG & \checkmark     & \checkmark     & \checkmark     &  \\
    LP-MusicCaps \cite{doh2023lp} &       & Whisper+BART-base & AAC   & \checkmark     & \checkmark     &       & \checkmark \\
    SongEditor \cite{yang2025songeditor} &       & RVQ+LM+DiT & TTM, ME  & \checkmark     & \checkmark     &       & \checkmark \\
    \midrule
    AudioLM \cite{Borsos2023audiolm} & Speech, Music & SoundStream+w2v-BERT & SC, PC & \checkmark     &       & \checkmark     &  \\
    \midrule
    AudioGEN \cite{kreuk2023audiogen} & \multirow{10}[2]{*}{Audio} & SoundStream+LSTM & TAC   & \checkmark     & \checkmark     & \checkmark     &  \\
    AudioLDM \cite{liu2023audioldm} &       & CLAP+LDM+VAE & TTA   & \checkmark     & \checkmark     & \checkmark     &  \\
    DualSpec \cite{zhao2025dualspec} &       & VAE+Flan-T5+LDM & TTSA   & \checkmark     & \checkmark     & \checkmark     &  \\
    ClipSep \cite{dong2023clipsep}&       & CLIP+U-Net & LASS  & \checkmark     & \checkmark     & \checkmark     &  \\
    AudioSep \cite{liu2024separate} &       & CLIP/CLAP+ResUNet & LASS, SE & \checkmark     & \checkmark     & \checkmark     &  \\
    FlowSep \cite{yuan2025flowsep}&       & FLAN-T5+VAE+BigVGAN & LASS  & \checkmark     & \checkmark     & \checkmark     &  \\
    TQ-SED \cite{yin2024TQSED} &       & CLAP+ResUNet+DPRNN & TSED  & \checkmark     & \checkmark     & \checkmark     &  \\
    WSTAG \cite{Xu2024TAG} &       & CRNN+CLAP & TAG  & \checkmark     & \checkmark     & \checkmark     &  \\
    EnCLAP \cite{kim2024enclap} &       & CLAP+BART+EnCodec & AAC   & \checkmark     &       &       & \checkmark \\
    DRCap \cite{li2024drcap} &       & CLAP+SLAM LLM & AAC   & \checkmark     &       &       & \checkmark \\
    \bottomrule
    \end{tabular}%
  \label{tab2:task-specific fine-tuning models}%
\end{table*}%

\subsubsection{Models for Discriminative Tasks}
Through contrastive learning with natural language supervision, audio-language pre-trained models can perform zero-shot audio classification and retrieval tasks within a unified framework, as shown in Fig. \ref{fig6:discriminative_models}. Although pre-trained models already have strong generalization capabilities, supervised fine-tuning on downstream datasets remains an important step to enhance task performance \cite{elizalde2023clap, wu2023large, elizalde2024CLAP2}. Audio classification refers to mapping input audio to a label, and the audio encoder from the pre-trained model can serve as a powerful audio pre-training model, used to train additional classifiers on specific datasets for fine-tuning. Retrieval can be divided into audio-to-text (A-T) and text-to-audio (T-A), aiming to filter corresponding samples of another modality from a pool based on the input, with classification considered as a special case of retrieval.

ATR research primarily focuses on enhancing system performance by integrating orthogonal SOTA works. 
For instance, CNN14-NetRVLAD \cite{lou2022contextatr} explores various audio features and sequence aggregation methods to improve audio-text alignment, utilizing the CNN14-based audio pretraining model PANNs and NetRVLAD pooling techniques to achieve significant improvements in bidirectional audio-text retrieval. 
EN-CA-IMC \cite{hu2023enatr} implements an audio enhancement strategy (EN) involving Gaussian noise, pitch adjustment, and time shift to reduce model overflow, combined with a co-attentive mechanism (CA) and intra-modal contrastive learning (IMC) between enhanced and original audio to capture richer audio features. 
CED-BERT-LE \cite{yan2024bridging} integrates a language enhancement strategy (LE) for effective cross-lingual retrieval and uses consistent ensemble distillation (CED) to handle variable-length audio segments, demonstrating proficiency across seven additional languages with minimal extra training data. 
Recently, the mini-batch Learning-to-match (m-LTM) framework \cite{luong2024revisiting} adapts the Learning-to-match approach using mini-batch subsampling and Mahalanobis-enhanced ground metrics, leveraging soft-matching from entropic optimal plans and Mahalanobis distance to learn a rich, expressive joint embedding space for audio and text modalities.

\subsubsection{Models for Generative Tasks}
Audio generation aims to enable the model to learn to generate audio that better conforms to textual conditions, facilitating the creation of personalized audio for applications such as augmented reality and virtual reality, game development, and video editing \cite{liu2023audioldm}.

A common method is to use a codec structure that converts continuous audio into discrete tokens for language modeling, ensuring high fidelity \cite{wu2024towards}. AudioGEN \cite{kreuk2023audiogen} is an pioneering audio language codec that uses data augmentation techniques, multi-stream modeling, and classifier-free guidance to improve adherence to text, addressing challenges such as distinguishing multiple sound sources, handling real-world recording conditions, and encoding high-fidelity audio. VALL-E \cite{wang2023VALLE}, VALL-E X \cite{zhang2023VALLEX}, MusicLM \cite{agostinelli2023musiclm} and SongEditor \cite{yang2025songeditor} perform well in speech and music respectively. SpeechX \cite{wang2024speechx} combines neural codec language modeling with task-specific prompts to achieve unified and scalable modeling and provides a consistent method for utilizing text input in speech enhancement and conversion tasks. Recent approaches further advance the development of audio generation technology by combining advanced generative models such as flow models and Diffusion. Flow-based models like Flow-TTS \cite{miao2020flow} improve the overall fluency and naturalness of generated speech by using reversible transformations to balance local accuracy and global coherence. Extending this approach, UniFlow-Audio \cite{xu2025uniflow} establishes a universal non-autoregressive framework for general audio generation, handling both time-aligned and non-time-aligned tasks via a dual-fusion mechanism that integrates features through temporal alignment and cross-attention. 
Seed-TTS \cite{anastassiou2024seedtts} is a foundational speech generation model that excels in speech in-context learning and achieves good performance in speaker similarity and naturalness.

\begin{figure}[ht]
    \centering
    \includegraphics[width=0.48\textwidth]{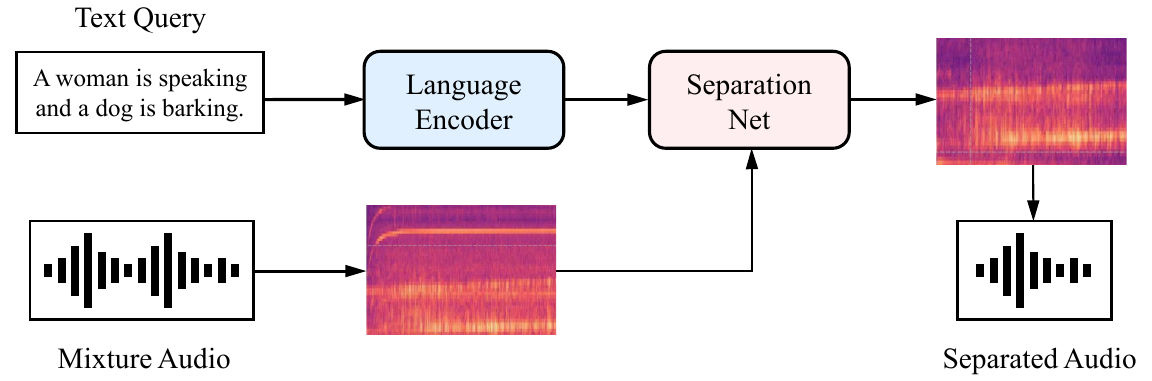}
    \caption{Illustration of language query-based audio separation model.}
    \label{fig7:audiosep}
\end{figure}

Audio source separation is a crucial audio generation task in enhancement applications. The language-queried audio source separation (LASS) task, which aims to isolate a target sound from a mixture based on natural language descriptions (e.g., ``a woman speaking"), typically requires tailored architectures when transferring ALMs. 
A representative approach is AudioSep \cite{liu2024separate}, which leverages a pre-trained ALM with aligned audio-text representations as its text query encoder. As illustrated in Fig. \ref{fig7:audiosep}, this encoder connects to a separation network and is fine-tuned on task-specific data, enabling source separation from free-form text queries and extending to tasks like speech enhancement. This demonstrates the potential of using natural language prompts to unify various audio enhancement tasks. Current research focuses on improving separation quality \cite{yuan2025flowsep} and model generalizability \cite{mahmud2024opensep}.

Recent works have explored spatial audio generation, advancing the field from single-channel to multi-channel and from non-directional to controllable formats. DualSpec \cite{zhao2025dualspec} is the first framework to directly generate spatial stereo audio with directional information from text descriptions, through a dual-branch design that integrates Mel-spectrograms and short-time Fourier transform spectrograms. SALM \cite{hu2025salm} aligns spatial audio with language via multi-modal contrastive learning. Its dual-branch encoder decomposes audio into semantic and spatial components, producing structured embeddings that enable zero-shot direction classification and text-based spatial audio editing. These technological advancements provide prospects for constructing more realistic audio datasets through synthetic data augmentation methods.

\begin{figure}[ht]
\centering
\includegraphics[width=0.48\textwidth]{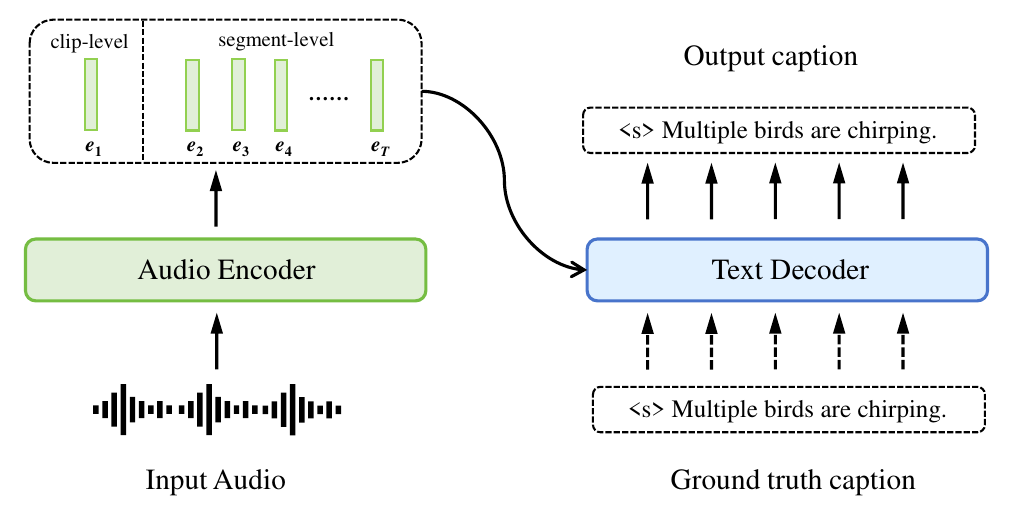}
\caption{Illustration of encoder-decoder language generative architecture.}
\label{fig8:AAC}
\end{figure}

Language generation aims to enable the model to perform audio-centric text prediction tasks, and fine-tuning can be conducted through language generative objectives. The two sub-tasks, audio understanding and language generation, are accomplished by the audio encoder and the text decoder, respectively. Recent work has also built upon this basic architecture by adding multiple encoders or decoders \cite{xu2023S2TT}.

Automated Audio Captioning (AAC) is an emerging task that involves recognizing environments, sound events, and their temporal relationships, then describes them in fluent sentences. The standard framework uses encoder-decoder architectures, with recent studies focusing on novel networks and training schemes for improvements\cite{xu2023aac_survey}.
The Audio Captioning Transformer (ACT) \cite{mei2021audio} is the first full Transformer architecture designed specifically for AAC, excelling in capturing global information and temporal relationships within audio signals. EnCLAP \cite{kim2024enclap} integrates two powerful acoustic representation models, EnCodec and CLAP, with the pre-trained language model BART to generate more accurate and contextually relevant captions. AutoCap \cite{haji2026taming} leverages extra guidance from metadata to enhance feature extraction, capturing nuanced acoustic information. LP-MusicCaps \cite{doh2023lp} addresses data scarcity in music captioning by using LLMs to generate descriptions from large-scale tag datasets.

\subsection{Multi-task Tuning and Cooperation}

In the early stages of audio processing, research efforts were primarily focused on breaking down complex tasks into simpler, independent audio tasks. Some studies \cite{deshmukh2023pengi, das2024speechverse} suggest that while these task-specific models were effective, they lacked understanding capabilities and failed to capture the connections between tasks, resulting in poor scalability and limited robustness when dealing with the inherent variability and complexity of audio signals. Some remarkable work has combined LLM in ALM achieving understanding and reasoning across diverse audio tasks. This convergence of large-scale language modeling with audio processing has opened up new possibilities to create more powerful, context-aware, and human-like audio understanding and reasoning systems.

\begin{table*}[htbp]
  \centering
  \caption{Summary of Popular Multi-task Audio-Language Models and Agent Systems}
    \begin{tabular}{cccccccc}
    \toprule
    \multirow{2}[3]{*}{Type} & \multirow{2}[3]{*}{Model} & \multicolumn{1}{c}{\multirow{2}[3]{*}{Domain}} & \multirow{2}[3]{*}{Architectures / Pre-trained models} & \multicolumn{2}{c}{Input} & \multicolumn{2}{c}{Output} \\
\cmidrule{5-8}          &       & \multicolumn{1}{c}{} &       & A     & T     & A     & T \\
    \midrule
    \multirow{5}[2]{*}{Codec} 
          & UniAudio \cite{yang2024uniaudio} & Audio & Tokenizers+Multi-scale Transformer & \checkmark     & \checkmark     & \checkmark     & \checkmark \\
\cmidrule{2-8}
          & SpeechX \cite{wang2024speechx} & \multirow{3}[1]{*}{Speech} & EnCodec+AR/NAR LM (VALL-E) & \checkmark     & \checkmark     & \checkmark     &  \\
          & VioLA \cite{wang2023viola} & \multicolumn{1}{c}{} & LSTM+AR LM & \checkmark     & \checkmark     & \checkmark     & \checkmark \\
          & LauraGPT \cite{du2023lauragpt} & \multicolumn{1}{c}{} & EnCodec+Qwen+Vododer & \checkmark     & \checkmark     & \checkmark     & \checkmark \\
          & Speech-XL \cite{sun2026speechxl} & \multicolumn{1}{c}{} & Whisper+Q-former+Qwen-2.5 & \checkmark & \checkmark &  & \checkmark \\
    \midrule
    \multirow{16}[4]{*}{LALM} & Pengi \cite{deshmukh2023pengi} & \multirow{8}[1]{*}{Audio} & CLAP+GPT-2 & \checkmark     & \checkmark     &       & \checkmark \\
          & LTU \cite{gong2024ltu}  & \multicolumn{1}{c}{} & AST+LLaMA & \checkmark     & \checkmark     &       & \checkmark \\
          & SALMONN \cite{tangsalmonn} & \multicolumn{1}{c}{} & Q-Former+Whisper+BEATs+Vicuna & \checkmark     & \checkmark     &       & \checkmark \\
          & Qwen-audio \cite{chu2023qwenaudio} & \multicolumn{1}{c}{} & Whisper+Qwen LM & \checkmark     & \checkmark     &       & \checkmark \\
          & Qwen-audio2 \cite{chu2024qwen2audio} & \multicolumn{1}{c}{} & Whisper+Qwen LM & \checkmark & \checkmark &       & \checkmark \\
          & Audio Flamingo \cite{kong2024audioflamingo} & \multicolumn{1}{c}{} & CLAP+OPT-IML-MAX & \checkmark     & \checkmark     &       & \checkmark \\
          & Audio Flamingo 2 \cite{ghosh2025audioflamingo2} & \multicolumn{1}{c}{} & AF-CLAP+XATTN-Dense+Qwen-2.5 & \checkmark     & \checkmark     &       & \checkmark \\
          & Audio Flamingo 3 \cite{goel2025flamingo3} & \multicolumn{1}{c}{} & AF-Whisper+Audio Adaptor+Qwen-2.5 & \checkmark     & \checkmark & \checkmark & \checkmark \\
          & GAMA \cite{ghosh2024gama} & \multicolumn{1}{c}{} & AST+Q-Former+LLaMa-2 & \checkmark     & \checkmark     &       & \checkmark \\
          & MiMo-Audio \cite{zhang2025mimo} & \multicolumn{1}{c}{} & MiMo-Audio-Tokenizer+MiMo-7B-Base & \checkmark & \checkmark & \checkmark & \checkmark \\
\cmidrule{2-8}          
          & SpeechGPT \cite{zhang2023speechgpt} & \multirow{8}[2]{*}{Speech} & HuBERT+LLaMA+HiFi-GAN & \checkmark     & \checkmark     & \checkmark     & \checkmark \\
          & SALM \cite{Chen2024SALM} & \multicolumn{1}{c}{} & Conformer+Megatron LLM & \checkmark     & \checkmark     &       & \checkmark \\
          & BLSP \cite{wang2023blsp} & \multicolumn{1}{c}{} & Whisper+Llama-2 & \checkmark     & \checkmark     & \checkmark     & \checkmark \\
          & SpeechT5 \cite{ao2021speecht5} & \multicolumn{1}{c}{} & wav2vec 2.0+Transformer+vocoders & \checkmark     & \checkmark     & \checkmark     & \checkmark \\
          & Voicebox \cite{NIPS2023Voicebox} & \multicolumn{1}{c}{} & ALiBi+VB-En/VB-Multi & \checkmark     & \checkmark     &       &  \\
          & SpeechVerse \cite{das2024speechverse} & \multicolumn{1}{c}{} & WavLM/Best-RQ/Others+Flan-T5 & \checkmark     & \checkmark     &       & \checkmark \\
          & Moshi \cite{defossez2024moshi} & \multicolumn{1}{c}{} & Mimi+Helium & \checkmark     & \checkmark     & \checkmark     & \checkmark \\
          & SHANKS \cite{chiang2025shanks} & \multicolumn{1}{c}{} & Qwen-omni & \checkmark     &      & \checkmark &  \\
\cmidrule{2-8}          
          & MusiLingo \cite{deng2024musilingo} & \multirow{2}[1]{*}{Music} & MERT+Vicuna & \checkmark     & \checkmark     &       & \checkmark \\
          & Llark \cite{gardner2023llark} & \multicolumn{1}{c}{} & Jukebox+Llama-2 & \checkmark     & \checkmark     &       & \checkmark \\
    \midrule
    \multirow{3}[2]{*}{Agent System} 
          & AudioGPT \cite{huang2024audiogpt} & Audio & GPT-3.5+Audio Foundation Models & \checkmark     & \checkmark     & \checkmark     & \checkmark \\
          \cmidrule{2-8}
          & SpeechAgents \cite{zhang2024speechagents} & Speech & SpeechGPT+LLaMA-2 & \checkmark     & \checkmark     & \checkmark     & \checkmark \\
          \cmidrule{2-8}
          & MusicAgents \cite{yu2023musicagent} & Music & LLMs+ Music Tools  & \checkmark     & \checkmark     & \checkmark     & \checkmark \\
    \bottomrule
    \end{tabular}%
  \label{tab3:multi-task models and systems}%
\end{table*}%

As listed in Table \ref{tab3:multi-task models and systems}, the popular ALMs can be categorized into two types: Multi-task Models (including codec-based models and LALMs) and Audio-Language Agent Systems.

\subsubsection{Multi-task Tuning Models}

\begin{figure}[ht]
\centering
\includegraphics[width=0.48\textwidth]{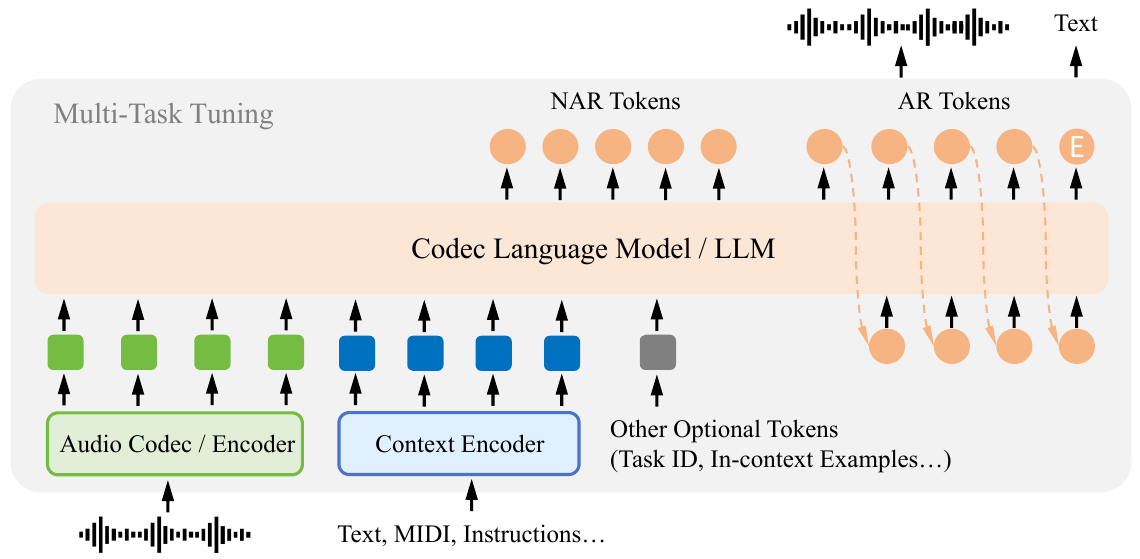}
\caption{Illustration of multi-task tuning.}
\label{fig:multi-task tuning}
\end{figure}

Multi-task tuning relies on a unified learning framework as illustrated in Fig. \ref{fig:multi-task tuning}, which can be achieved by multi-task sequence-to-sequence modeling, instruction tuning or in-context learning. 

Multi-task sequence-to-sequence modeling is a classic approach involves designing a multi-task sequence format for jointly represented tokens, which are widely adopted by codec-based ALMs. Codecs convert continuous audio into discrete tokens, which can easily be merged with text, MIDI, and other contextual tokens, and then passed to a language model to generate an output sequence that is further converted into the desired audio, text, or other outputs as needed for the task \cite{wang2023codec}. This codec-based pipeline based on discrete token has the advantage of high fidelity and is widely used in generative tasks. Moreover, its flexible encoding allows for more natural interaction between text and audio, enabling the use of a single model to perform ASR, TTS, S2TT, MT, and S2ST tasks, using context or ID to indicate the task to which a sample belongs \cite{wang2023viola}. Depending on the type of language model employed, codecs can be categorized into autoregressive (AR) and non-autoregressive (NAR) models. However, they may suffer from information loss due to the quantization of voice signals into discrete tokens, leading to significant performance degradation in some tasks compared to models that use continuous speech features \cite{du2023lauragpt}. 
Apart from codecs, Large Audio-Language Models (LALMs) aims to leverage the emergent understanding and reasoning capabilities of LLMs to build more generalized audio-centric models. These models typically employ an audio pre-trained model as the encoder, set up encoders for other modalities as needed, and add adaptive modules before connecting an LLM on top. However, pre-trained LLMs still lack sufficient cross-task generalization capabilities and require further post-training \cite{zhao2023LLMsurvey}. Multi-task tuning for LALMs can be similar to codecs, adopting a unified sequence format, or achieved through instruction tuning or in-context learning. 
VioLA \cite{wang2023viola} is a Transformer-based autoregressive decoder that unifies speech-text cross-modal tasks. By converting speech to discrete tokens and integrating task/language IDs, it effectively handles multiple languages and tasks, demonstrating excellent performance in both single-modal and cross-modal settings.
LauraGPT \cite{du2023lauragpt} operates based on a unified task expression (input embeddings, task ID, output tokens) to handle multiple tasks. Its modular and flexible design allows the model to perform complex tasks by combining sub-tasks such as S2ST.
UniAudio \cite{yang2024uniaudio} is an advanced codec-based ALM that supports 11 diverse audio generation tasks. It tokenizes both target audio and various condition modalities, concatenating them into unified [conditions, target] sequences. The system employs a multi-scale Transformer architecture for next-token prediction, efficiently handling the extended sequences from neural codec tokenization.
Pengi \cite{deshmukh2023pengi} frames all audio tasks as text-generation tasks, allowing for comprehensive audio understanding and generation. Pengi demonstrates remarkable capabilities across zero-shot tasks in general audio including environmental sound, speech, and music tasks, and achieves good performance in close-ended and open-ended audio tasks.
Qwen-audio \cite{chu2023qwenaudio} employs a multi-task training framework that uses hierarchical tags to condition the decoder, enabling effective knowledge sharing without interference. This allows the model to handle diverse audio types and tasks, including ASR, AAC, and audio question answering.
SHANKS \cite{chiang2025shanks} and STITCH \cite{chiang2025stitch} exemplify this through specialized training sequences that interleave input, reasoning, and output. SHANKS interleaves user speech chunks with internal "thinking" tokens to learn incremental reasoning for low-latency intervention. STITCH alternates between reasoning and speech output tokens, enabling concurrent ``thinking" and ``speaking".

Instruction tuning is another one of the current mainstream methods, which aims to provide multiple audio tasks with unified instance construction formats using natural language. This allows a single model to handle multiple tasks without explicit fine-tuning on specific tasks but by enhancing its instruction-following capability \cite{liu2024instructiontuning}.
SpeechGPT \cite{zhang2023speechgpt} incorporates discrete speech representations into a large language model framework, enabling intrinsic cross-modal conversational abilities. 
SpeechGPT is built upon SpeechInstruct, a pioneering large-scale cross-modal speech instruction dataset, and employs a sophisticated three-stage training strategy: modality-adaptation pre-training, cross-modal instruction fine-tuning, and chain-of-modality instruction fine-tuning. This approach allows SpeechGPT to perceive and generate multi-modal content effectively, demonstrating impressive capabilities in following cross-modal human instructions and showcasing the potential for handling multiple modalities within a single model.
LLark \cite{gardner2023llark} and MusiLingo \cite{deng2024musilingo} are instruction-tuned multimodal model designed specifically for music tasks, from classification and regression to more complex captioning and reasoning challenges and has demonstrated impressive capabilities in zero-shot tasks. 
SALMONN \cite{tangsalmonn} employs a three-stage training process: pre-training, instruction tuning, and an additional activation tuning stage designed to mitigate overfitting to speech recognition and audio captioning tasks. It demonstrates strong zero-shot and downstream performance across environmental sound, speech, and music, while introducing two novel tasks: audio-based storytelling and speech-audio co-reasoning.
LTU \cite{gong2024ltu} integrates LLM comprehension with audio foundation models for listening, thinking, and understanding. Trained on the OpenAQA-5M dataset consisting of closed-ended and open-ended diverse (audio, question, answer) tuples, using a perception-to-understanding curriculum. It excels in conventional tasks such as classification and captioning, as well as emerging audio reasoning and compression abilities that are absent in existing audio models.
GAMA \cite{ghosh2024gama} integrates multiple types of audio representations using a custom Audio Q-Former and a multi-layer aggregator, capturing diverse aspects from semantic generalization to surface-level audio properties. They introduced CompA-R and CompA-R-test, a human-labeled dataset for open-ended audio question answering requiring complex reasoning. Fine-tuned on CompA-R, GAMA demonstrates strong performance in complex reasoning and instruction-following tasks.

In-context learning enables LLMs to quickly adapt to specific tasks using a few examples, allowing them to predict unseen input labels without additional parameter updates. In the transfer of ALMs, the goal is to learn and enhance vocabulary from the context of prompt texts without performing backpropagation \cite{Chen2024SALM}, thereby unlocking the capabilities of frozen LLMs for a wide range of audio understanding and generation tasks \cite{yang2024uniaudio1.5}.
Voicebox \cite{NIPS2023Voicebox} is a non-autoregressive flow-matching model trained to infill speech based on audio context and text, utilizing over 50,000 hours of unfiltered or enhanced speech data for training. Voicebox can perform various tasks through in-context learning, such as mono or cross-lingual zero-shot TTS, noise removal, content editing, style conversion, and diverse sample generation, achieving high fidelity and efficiency in speech synthesis and editing.
SALM \cite{Chen2024SALM} equipped with multitask and in-context learning capabilities. SALM achieving comparable performance to task-specific models through instruction tuning for ASR and AST. Additionally, a speech-supervised in-context training method is proposed to bridge the gap between LLM training and downstream speech tasks, further enhancing the in-context learning ability of speech-to-text models.
Audio Flamingo \cite{kong2024audioflamingo} achieves leading performance on various audio understanding tasks without requiring task-specific fine-tuning, supporting instruction tuning for robust multi-turn dialogue and demonstrating rapid adaptation through in-context learning. Building on this foundation, Audio Flamingo 2 extends these capabilities to long-form audio understanding and introduces the LongAudioBench benchmark \cite{ghosh2025audioflamingo2}. The recently developed Audio Flamingo 3 \cite{goel2025flamingo3} further expands both training data and learning curricula, achieving breakthrough advancements in reasoning and understanding across speech, sound, and music domains.


\subsubsection{Audio-Language Agent Systems}

\begin{figure}[h]
\centering
\includegraphics[width=0.42\textwidth]{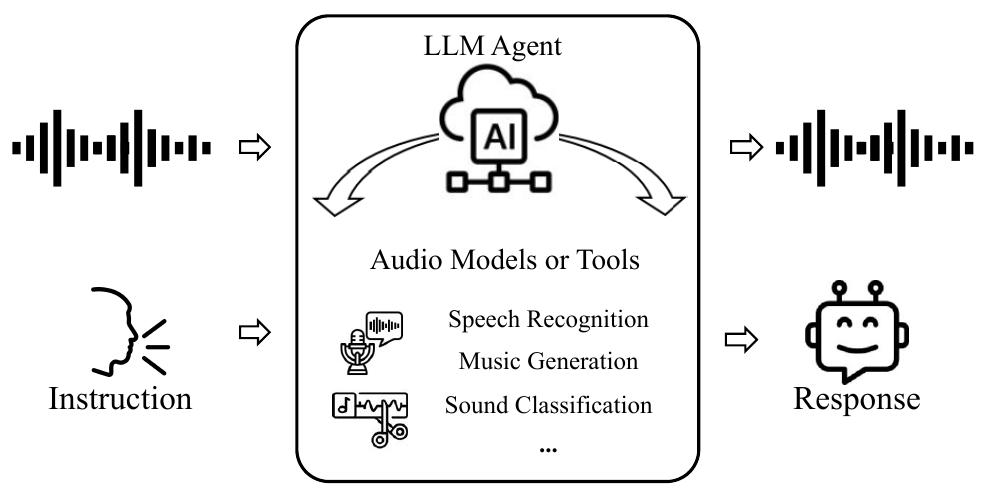}
\caption{Illustration of audio-language agent systems.}
\label{fig:agent system}
\end{figure}

Agent Systems utilize LLMs as agents to transform human instructions, achieving task transfer in downstream tasks by cooperating different foundational pre-trained models, as shown in Fig. \ref{fig:agent system}. By leveraging the reasoning capabilities of LLMs, these agent systems are designed to invoke and collaborate with a diverse array of audio models and tools to accomplish complex tasks initiated by humans.

SpeechAgents \cite{zhang2024speechagents} is a multimodal and multi-agent system that proposes a multi-agent tuning technique. The agents collaborate in a distributed and parallel manner to make decisions and execute actions, thereby significantly improving work efficiency and effectiveness. With excellent scalability, SpeechAgents is particularly suitable for tasks such as drama creation and audio novel generation.
MusicAgent \cite{yu2023musicagent} is a system designed to integrate and organize various music processing tasks. Driven by LLMs, it automatically decomposes user requests into multiple sub-tasks and invokes the appropriate music tools to fulfill these requirements. MusicAgent can handle a wide range of music processing tasks and aims to bridge the gap between tools from different sources by unifying data formats (text, MIDI, ABC notation, audio), enabling seamless collaboration among tools on different platforms.
AudioGPT \cite{huang2024audiogpt} is an innovative multimodal system that integrates ChatGPT with audio foundation models and input/output interfaces to process complex audio information and conduct spoken conversations. AudioGPT demonstrates proficiency in various audio-related tasks, including speech processing, music analysis and generation, and sound effect understanding and creation.

Notably, while Tables \ref{tab2:task-specific fine-tuning models} and \ref{tab3:multi-task models and systems} document architectural relationships between pre-trained models and downstream tasks, systematic evaluations comparing pre-training methods for specific applications remain limited. This gap stems from rapidly evolving task paradigms and ALMs' multi-component complexity, which complicates generalizable model selection \cite{elizalde2024CLAP2}. (For representative evaluations on commonly assessed tasks like AC and ATR, see Section \ref{eval results}.) We therefore recommend experimental evaluation within specific application contexts to identify suitable pre-trained models.

Beyond supervised fine-tuning (SFT) paradigms discussed above, reinforcement learning (RL) and preference optimization are emerging as new directions for ALM downstream transfer. RL enables learning from evaluative feedback via cumulative reward maximization, showing potential to surpass SFT in audio reasoning \cite{li2025RL, wu2025audiothinker}. Direct preference optimization (DPO) offers a lightweight alternative, learning from comparative judgments without explicit reward modeling \cite{huang2025step}, and proves effective for subjective qualities like naturalness and musicality \cite{broukhim2025prefersurvey}.

\section{Datasets and Benchmarks} \label{section: Data}
\begin{figure}[htbp]
  \includegraphics[width=0.49\textwidth]{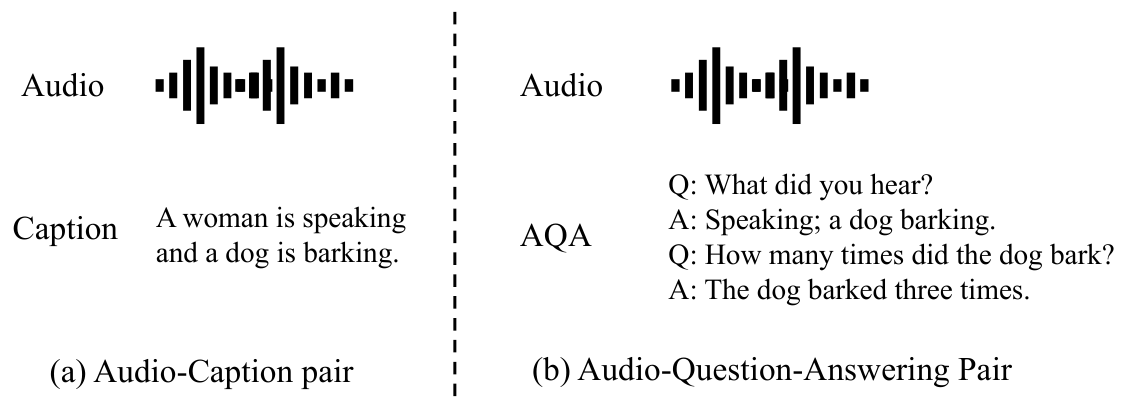}
  \caption{Illustration of two types of audio-language datasets.}
  \label{fig11: datasets}
\end{figure}

In this section, we review the audio-language datasets, which primarily include audio-text paired datasets and audio question answering datasets (shown in Fig. \ref{fig11: datasets}).

\subsection{Audio-Text Paired Datasets}
Audio-text paired datasets contain audio recordings paired with corresponding text, including captions (describing audio content), transcriptions (for speech), or translations (transcriptions in other languages). Natural language descriptions capture multiple audio events more freely and in detail, making audio-caption datasets particularly valuable for representation pre-training. Transcriptions and translations are typically used for task-specific model training. Table \ref{tab4:paired datasets} summarizes commonly used audio-text paired datasets for ALM training.

\begin{table*}[htbp]
  \centering
  \caption{Summary of audio-text paired datasets.}
    \begin{tabular}{ccccc}
    \toprule 
    Datasets & Domain & Audio Source & Pairs & Text Source \\
    \midrule
    AudioCaps \cite{kim2019audiocaps} & \multirow{14}[1]{*}{Audio} & AudioSet & 57K   & Human \\
    Sound-VECaps \cite{yuan2025soundvecaps} &       & AudioSet & 1.6M  & Generated \\
    AudioGrounding \cite{xu2021AudioGrounding} &       & AudioSet & 5K    & Generated \\
    AudioSetCaps \cite{bai2025audiosetcaps} &       & AudioSet, Youtube-8M, VGGSound & 6M  & Generated \\
    LAION-Audio-630K \cite{wu2023large} &       & AudioSet, FreeSound, BBC Sound Effects, etc. & 630K  &  -\\
    WavCaps \cite{mei2024wavcaps} &       & AudioSet, FreeSound, BBC Sound Effects, SoundBible & 400K & Generated \\
    Auto-ACD \cite{sun2024auto} &       & AudioSet, VGGSound & 1.5M  & Generated \\
    Clotho \cite{drossos2020clotho} &       & FreeSound & 30K   & Human \\
    AudioDiffCaps \cite{Takeuchi2023AudioDiffCaps} &       & FreeSound & 29K   & Generated \\
    augmented FSD50K \cite{Liu2024decasesep} &       & FreeSound & 51K   & Online \\
    WavText5k \cite{deshmukh2022wavtext} &       & BigSoundBank, SoundBible & 5K    & Online \\
    SoundDescs \cite{koepke2022benchmarks} &       & BBC Sound Fffects & 33K   & Online \\
    MACS \cite{Martin2021bMACS} &       & TAU Urban Acoustic Scenes  & 17K & Human \\
    AudioCaption \cite{wu2019audio} &       & Internal Collected & 29K   & Human \\
    \cmidrule{2-5}
    EMOSpeech \cite{bian2024emospeech} & \multirow{2}[0]{*}{Speech} & Internal Collected & 31K   & Generated+Human \\
    SpeechCraf \cite{jin2024speechcraft} &       & AISHELL-3, Zhvoice, GigasSpeech-m, LibriTTs-R & 2.1M  & Generated \\
    \cmidrule{2-5}
    MusicCaps \cite{agostinelli2023musiclm} & Music & AudioSet & 5.5K  & Human \\
    \cmidrule{2-5}
    AnimalSpeak \cite{robinson2024bioacoustics} & Bioacoustic & iNaturalist, Xeno-Canto, Animal Sound Archive, etc. & 1.1M  & Generated \\
    \bottomrule
    \end{tabular}%
    \begin{tablenotes}
        \item[] AudioSet and VGGSound are sourced from YouTube.
    \end{tablenotes}
  \label{tab4:paired datasets}%
\end{table*}%

\subsection{Audio Question Answering Datasets}
Audio Question Answering Datasets are designed for training and evaluating models that can answer questions about audio content. These datasets are structured as triplets of audio, question, and answer. Initially, such datasets were specifically created for Audio Question Answering (AQA) tasks, constructing diverse questions to train models on how to answer similar queries. In recent years, AQA tasks have been increasingly utilized as a proxy task for model learning perception and reasoning about audio \cite{fayek2020temporal}. They can serve as a format that integrates multiple audio tasks \cite{gong2024ltu, Kuan2025ICASSP} and can also be used to guide models in better following human instructions \cite{zhang2023speechgpt}. We have listed several popular open-source QA datasets in Table \ref{tab5:QA datasets}.
Furthermore, data annotation is shifting from manual labeling toward synthetic and weakly-supervised approaches, including LLM-generated audio captions and self-supervised pseudo-labels (see “Text source” columns in Tables \ref{tab4:paired datasets} and \ref{tab5:QA datasets}, where “generated” denotes model-produced text). For detailed dataset analyses, we refer readers to dedicated audio-language dataset surveys \cite{wijngaard2024ald_survey}.

\begin{table*}[htbp]
  \centering
  \caption{Summary of Audio Question Answering Datasets.}
    \begin{tabular}{cccccc}
    \toprule
    \multirow{2}[4]{*}{Datasets} & \multirow{2}[4]{*}{Audio Source} & \multirow{2}[4]{*}{Pairs} & \multicolumn{3}{c}{Text Source} \\
\cmidrule{4-6}          &       &       & Template & Generated & Human \\
    \midrule
    DAQA \cite{fayek2020temporal} & AudioSet & 600K  & \checkmark     &       & \checkmark \\
    AudioSetCaps-QA \cite{bai2025audiosetcaps} & AudioSet, Youtube-8M, VGGSound & 18.4M  & \checkmark     &      &   \\
    SpeechInstruct \cite{zhang2023speechgpt} & Gigaspeech, Common Voice, LibriSpeech & 9M    & \checkmark     & \checkmark     &   \\
    ClothoAQA \cite{lipping2022clotho} & \multirow{2}[0]{*}{FreeSound} & 36K   &       &       & \checkmark     \\
    mClothoAQA \cite{behera23_interspeech} &       & 287K  &       & \checkmark     & \checkmark  \\
    OpenAQA-5M \cite{gong2024ltu} & AudioSet, VGGSound, Freesound, Sound Bible, etc. & 5.6M &       & \checkmark     & \checkmark  \\
    CompA-R \cite{ghosh2024compa} & AudioSet & 200K &       & \checkmark     & \checkmark    \\
    Audio Dialogues \cite{goel2024Dialogues} & AudioSet & 163.8K &       & \checkmark     &    \\
    MusicQA \cite{liu2024music} & AudioSet,TagATune  & 70K   &       &       & \checkmark   \\
    AudioDER \cite{geng2026audioder} & AudioSet, VGGSound, Freesound, LibriSpeech, etc.  & 191K &       & \checkmark & \checkmark   \\
    \bottomrule
    \end{tabular}%
  \label{tab5:QA datasets}%
\end{table*}%

\subsection{Benchmarks}
Evaluating ALMs involves integrating them with specific audio tasks and datasets, leading to variations in standards and makes model comparison unfair. To solve this, several benchmarks have been created, which not only release test datasets but also establish corresponding evaluation strategies. These benchmarks ensure strict and consistent evaluation of ALMs, enhancing scientific reliability.

Benchmarks for ALMs fall into three categories: task-specific, cross-task, and audio instruction. Task-specific benchmarks such as SoundDescs \cite{koepke2022benchmarks}, CompA \cite{ghosh2024compa}, and ADU-Bench \cite{gao2024ADUbench} evaluate performance on defined tasks including audio-text retrieval and audio dialogue. Cross-task benchmarks like ARCH \cite{la2024ARCH}, MMAU \cite{sakshi2025mmau}, MMAR \cite{ma2026mmar}, SSEU-bench \cite{yin2025SSEUbench} and AQUA-bench \cite{kuan2026aqua} assess model generalization across domains, requiring diverse training data. This category also includes speech-related cross-task benchmarks \cite{yang2021superb, Huang2024Dynamic-Superb, evain2021LeBenchmark, Ghosh2022LAPE, chen2024voicebench} and music-specific ones \cite{melechovsky2024mustango, weck2024muchomusic}. Audio instruction benchmarks including AIR-Bench \cite{yang2024air} and AudioBench \cite{wang2025audiobench} test LALMs' audio understanding and instruction-following abilities through unified conversational formats. Recent benchmark studies have begun exploring additional dimensions to measure ALMs' capabilities, with Audio Jailbreak \cite{song2025ajailbreak} and LongAudioBench \cite{ghosh2025audioflamingo2} focusing on adversarial robustness and long-audio processing capabilities, respectively.

These benchmarks aim to improve model performance and ensure fair comparison across different scenarios. The baseline performances on these benchmarks reflect the progress of model research and help identify current limitations to guide future studies, thereby advancing the field of audio-language learning and developing more powerful models.

\section{Evaluations}
\label{section: Evaluations}


\subsection{Evaluation Methods}
\subsubsection{Zero-Shot Evaluation} This method evaluates ALMs' open-set retrieval through similarity measurement between audio and text embeddings in a shared space. By treating labels as specialized language, it extends naturally to classification tasks on standard datasets like ESC50 \cite{piczak2015esc50}.

\subsubsection{Linear Probe Evaluation} It evaluates audio representation quality by training a linear classifier (e.g., multilayer perceptron) on downstream tasks with frozen ALM features. While not optimal for specific tasks, this setup minimizes variables for fair representation comparison, commonly used in audio classification.

\subsubsection{Supervised Fine-tune Evaluation} It examines model generalization by fine-tuning the entire model or its components on downstream tasks. The audio encoder, along with a task-specific head, is optimized end-to-end, and performance is validated on test sets with comparison to state-of-the-art (SOTA) models. It comprehensively assesses model adaptability by supporting fine-tuning across a wide range of tasks, such as classification, retrieval, and captioning \cite{drossos2020clotho, kim2019audiocaps} and so on.

\subsubsection{Instruction-following Evaluation} It measures the ability to understand and accurately follow human instructions, indicating task generalization in LALMs. This evaluation method can be considered a special type of zero-shot evaluation or supervised fine-tuning evaluation, depending on whether instruction tuning is performed. Recently, evaluation of general LALMs has shifted toward unified-format benchmarks \cite{sakshi2025mmau, ma2026mmar} that require instruction following for question answering, thus assessing multi-task performance.

\subsection{Evaluation Results}
\label{eval results}
Table \ref{tab:zs_lp} summarizes zero-shot and linear probe results (mAP: mean Average Precision; Acc: Accuracy) on common audio classification tasks, widely used for evaluating contrastive learning representations and LALM benchmarks. Results confirm contrastive learning enables open-set capability, while adaptive training through a simple linear classifier significantly improves downstream task performance. Table \ref{tab:ft_atr} shows supervised fine-tuning results on ATR, where R@k (Recall at rank k) reflects task adaptation capability. For instruction-following evaluation spanning diverse tasks and settings, readers are referred to established benchmarks \cite{yang2024air, wang2025audiobench} for systematic LALM comparisons.

Notably, the field faces reproducibility and fairness challenges in model evaluation. As systematically revealed in \cite{yang2025evaluationsurvey}, YouTube-sourced datasets (e.g., AudioSet, AudioCaps) suffer from link rot, leading to inconsistent data compositions across studies and affecting result comparability. Additionally, data leakage is prevalent—unintended audio overlaps between different datasets (e.g., WavCaps and Clotho/AudioCaps) can inflate evaluation metrics if left unaddressed. Therefore, interpreting existing benchmark results requires accounting for these inconsistencies in evaluation setups. Future research should establish standardized data usage protocols and stricter de-duplication pipelines to ensure evaluation fairness and result reproducibility.

\begin{table*}[htbp]
  \centering
  \caption{Zero-Shot and Linear Probe Evaluation of ALMs on Audio Classification Tasks (\%).}
    \begin{tabular}{lcccccccc}
    \toprule
    Dataset & \multicolumn{1}{c}{AudioSet \cite{gemmeke2017audioset}} & \multicolumn{1}{c}{FSD50K \cite{fonseca2022FSD50K}} & \multicolumn{1}{c}{ESC50 \cite{piczak2015esc50}} & \multicolumn{1}{c}{US8K \cite{Mesaros2018urban}} & \multicolumn{1}{c}{CRM-D \cite{cao2014crema}} & \multicolumn{1}{c}{GTZAN \cite{sturm2013gtzan}} & \multicolumn{1}{c}{NS \cite{engel2017ns}} & \multicolumn{1}{c}{VGG \cite{Chen20VGGSound}} \\
    \midrule
    Metric & \multicolumn{1}{c}{mAP} & \multicolumn{1}{c}{mAP} & \multicolumn{1}{c}{Acc} & \multicolumn{1}{c}{Acc} & \multicolumn{1}{c}{Acc} & \multicolumn{1}{l}{Acc} & \multicolumn{1}{c}{Acc} & \multicolumn{1}{c}{mAP} \\
    \midrule
    \textbf{Zero-Shot} &       &       &       &       &       &       &       &  \\
    MS-CLAP \cite{elizalde2023clap} & 5.8   & 30.2  & 82.6  & 75.3  & 22.8  & 29.0  & 21.4  & - \\
    LAION-CLAP \cite{wu2023large} & - & 45.9  & 91.0  & 77.0  & \underline{23.1}  & 47.2  & 35.3  & 30.0  \\
    MS-CLAP V2 \cite{elizalde2024CLAP2} & 10.2  & \underline{48.5}  & \underline{93.9}  & \textbf{82.3}  & \textbf{30.0}  & 58.4  & \textbf{58.1}  & - \\
    WavCaps \cite{mei2024wavcaps} & \underline{19.6}  & \textbf{53.0}  & \textbf{94.8}  & \underline{81.4}  & 19.9  & 45.5  & 27.7  & 28.9  \\
    BLAT \cite{xu2023blat} & 10.5  & 31.3  & 80.6  & 77.3  & - & - & - & 14.9  \\
    Pengi \cite{deshmukh2023pengi} & 16.4  & 46.8  & 92.0  & 71.9  & 18.5  & 35.3  & \underline{50.1}  & - \\
    LTU \cite{gong2024ltu} & 18.7  & 46.3  & 83.1  & - & - & 50.3  & - & - \\
    M2D-CLAP \cite{niizumi2024m2dclap} & \textbf{27.2}  & 40.8  & 75.5  & 72.4  & 17.7  & \textbf{75.2}  & 23.4  & - \\
    Cacophony \cite{zhu2024cacophony} & - & - & 93.4 & 77.1 & - & - & - & 27.1  \\
    GLAP \cite{dinkel2025glap} & - & 40.9  & 88.8  & 78.9  & 20.5  & \underline{69.6}  & 31.3  & - \\
    \midrule
    \midrule
    \textbf{Linear Probe} &       &       &       &       &       &       &       &  \\
    MS-CLAP \cite{elizalde2023clap} & - & - & 93.8  & 84.2  & 54.4  & 79.3  & 68.2  & - \\
    LAION-CLAP \cite{wu2023large} & - & - & \underline{97.3}  & 86.9  & 54.6  & 84.3  & 72.2  & - \\
    MS-CLAP V2 \cite{elizalde2024CLAP2} & - & - & \textbf{97.7}  & \underline{88.4}  & \underline{62.5}  & 82.3  & \textbf{80.5}  & - \\
    WavCaps \cite{mei2024wavcaps} & - & - & 97.2  & 63.6  & 58.6  & 80.2  & 74.4  & - \\
    BLAT \cite{xu2023blat} & 38.7  & 32.4  & 95.8  & 85.7  & - & - & - & 42.9  \\
    Pengi \cite{deshmukh2023pengi} & - & - & 89.2  & - & 50.6  & 80.0  & - & - \\
    M2D-CLAP \cite{niizumi2024m2dclap} & - & - & 96.3  & \textbf{88.8}  & \textbf{73.4}  & 84.1  & 78.0  & - \\
    \bottomrule
    \end{tabular}%
    \begin{tablenotes}
    \item[] Results are sourced from \cite{niizumi2024m2dclap, zhu2024cacophony, xu2023blat, dinkel2025glap}; the best and second-best scores are in bold and underline, respectively.
    \end{tablenotes}
  \label{tab:zs_lp}%
\end{table*}%

\begin{table*}[htbp]
  \centering
  \caption{Supervised Fine-tune Evaluation of ALMs on Audio-Text Retrieval Tasks (\%).}
    \begin{tabular}{lcccccccccccc}
    \toprule
    Dataset & \multicolumn{6}{c}{AudioCaps \cite{kim2019audiocaps}}                 & \multicolumn{6}{c}{Clotho \cite{drossos2020clotho}} \\
    \midrule
    Input-Output & \multicolumn{3}{c}{Text-to-Audio} & \multicolumn{3}{c}{Audio-to-Text} & \multicolumn{3}{c}{Text-to-Audio} & \multicolumn{3}{c}{Audio-to-Text} \\
    \midrule
    Metric & R@1 & R@5 & R@10 & R@1 & R@5 & R@10 & R@1 & R@5 & R@10  & R@1 & R@5 & R@10 \\
    \midrule
    MS-CLAP \cite{elizalde2023clap} & 33.5  & 70.4  & 80.2  & 47.8  & 80.2  & 90.7  & 16.2  & 39.6  & 51.4  & 23.6  & 46.7  & 60.3 \\
    LAION-CLAP \cite{wu2023large} & 35.1  & 71.8  & 83.9  & 45.8  & 80.9  & 91.6  & 18.2  & 42.5  & 54.4  & \underline{25.7} & 51.5  & 63.4 \\
    MS-CLAP V2 \cite{elizalde2024CLAP2} & 35.6  & - & - & 42.5 & 15.7 & - & - & - & - & 22.9  & - & - \\
    FLAP \cite{yeh2023flap}  & \underline{41.5}  & \textbf{75.5}  & 86.0 & 53.0 & \textbf{84.1} & \textbf{92.6} & \textbf{20.3}  & \textbf{46.5}  & \textbf{58.8}  & 25.5  & \underline{53.4}  & \textbf{67.9} \\
    WavCaps \cite{mei2024wavcaps} & 39.7  & 74.5  & 86.1  & 51.7  & 82.3  & 90.6  & 19.5  & 45.2  & 58.2  & 23.4  & 50.9  & 63.4 \\
    Cacophony \cite{zhu2024cacophony} & 41.0 & - & \underline{86.4}  & \textbf{55.3}  & \underline{83.6}  & \underline{92.4}  & \underline{20.2} & \underline{45.9}  & \textbf{58.8}  & \textbf{26.5}  & \textbf{54.1}  & \underline{67.3} \\
    BLAT \cite{xu2023blat} & 38.2 & - & 85.1  & 47.5  & - & 87.6  & 13.7  & - & 48.9  & 17.9  & - & 50.9 \\
    AutoACD \cite{sun2024auto} & 39.5  & - & 85.4  & 53.7  & - & 91.7  & 15.3  & - & 52.1  & 17.7  & - & 52.6 \\
    T-CLAP \cite{yuan2024tclap} & 39.7  & \underline{74.6}  & \textbf{86.9}  & 49.8  & 82.5  & 91.9  & 17.3  & 39.9  & 53.6  & 21.8  & 44.9  & 57.4 \\
    GLAP \cite{dinkel2025glap}  & \textbf{41.7}  & - & 86.1  & \underline{54.4}  & - & 91.1  & 19.4  & - & 58.3  & 21.8  & - & 61.5 \\
    \bottomrule
    \end{tabular}%
    \begin{tablenotes}
    \item[] Results are sourced from \cite{niizumi2024m2dclap, zhu2024cacophony, xu2023blat, dinkel2025glap, yuan2024tclap}; the best and second-best scores are in bold and underline, respectively.
    \end{tablenotes}
  \label{tab:ft_atr}%
\end{table*}%

\section{Core Limitations and Concerns}\label{section: Limitations and Concerns}



\subsection{Security Issues}

At the input level, although models demonstrate reasonable robustness to minor perturbations such as background noise \cite{achiam2023gpt4}, they remain vulnerable to adversarial perturbations, spoofed audio, and prompt injection attacks \cite{kang2025advwave, song2025ajailbreak, yang2025AchillesHeel}. At the model level, fundamental limitations such as hallucination\cite{kuan2024hallucination, arora2025landscape}, incorrect temporal reasoning \cite{ghosh2024compa}, and over-reliance on textual priors \cite{ma2026mmar}. These input-level attacks and model-level flaws may further lead the model to generate harmful or misleading content at the output level, posing direct risks to users \cite{jin2026almguard}. To address the above risks, potential solutions include input filtering and prompt reminder mechanisms \cite{xie2023defending}, adversarial training \cite{song2025ajailbreak}, as well as additional deployment measures such as output filtering and human-in-the-loop verification.

\subsection{Privacy Risks}

ALMs introduce privacy risks by processing raw audio directly. Unlike cascaded pipelines, end-to-end models retain rich acoustic information (e.g., intonation, speaker identity), potentially enabling unauthorized user tracking via implicit voiceprint storage and inference of sensitive attributes like age and emotion \cite{he2025model}. Environmental audio further risks exposing location, activities, or habits through soundscape monitoring, enabling covert large-scale surveillance beyond legal safeguards. Mitigation spans model-level techniques including safety alignment and in-context unlearning, and data-level protection such as anti-eavesdropping jamming \cite{wang2025man}.

\subsection{Bias}

Web-scale training inherits and amplifies societal biases through imbalanced concept representation and spurious correlations \cite{lin2024emobias, slaughter2023pre}, manifesting across multiple dimensions. In terms of demographic variations, SER models may associate emotions with gender, while ASR performance degrades for underrepresented subgroups such as specific age or gender groups \cite{koudounas2023subgroup}. Linguistic bias is also prevalent: the dominance of high-resource languages (e.g., English) in training data leads to significant performance disparities for low-resource languages, accents, and dialects \cite{Peng2025SLLMsurvey, arora2025landscape}. Furthermore, domain-specific context gaps, i.e., the mismatch between general-purpose training data and specialized domains such as healthcare and legal, lead to significant performance degradation in particular domains. Mitigations include data augmentation \cite{zhang2023dataaug}, focused data collection \cite{koudounas2024prioritizing}, and fairness-aware losses \cite{zhang2022mitigating}. Most existing research operates on predefined subgroups such as race or valence, motivating alternative paradigms such as contrastive learning for latent representations \cite{koudounas2024contrastivebias} and confidence-oriented debiasing \cite{tsai2025co}.

\subsection{Training Costs}
Scaling data and models creates significant accessibility and sustainability barriers. This is exemplified by contrastive pre-training, which requires large batches (e.g., FLAP used 72×64 GPUs for a 4608 batch size \cite{yeh2023flap}). Although parameter-efficient methods like LoRA \cite{hu2022lora} reduce LLM fine-tuning costs, LALMs still need extensive training to bridge modality gaps \cite{gong2024ltu}. Representative examples include LTU (5.6M samples) and Audio Flamingo (5.9M samples; 8 A100s) \cite{gong2024ltu, kong2024audioflamingo}. However, computational requirements for adapting ALMs vary significantly across downstream tasks. For example, TQ-SED \cite{yin2024TQSED} fine-tunes only 0.25M parameters, demonstrating the potential for broader ALM deployment.

\section{Future Research Directions} \label{section: Future directions}

\subsection{Towards Efficient and Scalable ALMs}
Addressing the high computational cost of ALMs necessitates a shift towards more efficient and scalable learning methods. Promising directions encompass: \textbf{Parameter-Efficient Architectures} which leverage techniques such as model compression via distillation \cite{liu2024audiolcm, paissan2024tinyclap} for compact edge models and adapter-based tuning for adaptation with minimal parameters; \textbf{Data-Efficient Training} utilizing active or curriculum learning \cite{wan2023activesurvey} to maximize limited labeled data utility; and \textbf{Lifelong Adaptation} via continual learning \cite{yang2024aqacontinual} that enable knowledge accumulation scalable to non-stationary data streams without catastrophic forgetting, thus avoiding expensive retraining cycles.

\subsection{Strengthening Security Safeguards}
As ALM generation capabilities advance, so do the risks of malicious use, particularly through deepfake audio creation. Future research must prioritize developing robust detection frameworks \cite{li2023voiceguard} that can identify synthetic audio across diverse generation methods. Concurrently, applications such as voice assistants necessitate the integration of encryption, authentication, and access control technologies to provide technical safeguards against misuse \cite{Simon2023AssistantSecurity}. Additionally, exploring adversarial training techniques and developing certified defenses against audio manipulation attacks will be crucial for building secure and trustworthy ALM systems.

\subsection{Mitigating Algorithmic and Ethical Bias}
Algorithmic and ethical bias in ALMs extends beyond simple demographic subgroups, manifesting in complex and systemic ways. For example, linguistic bias is particularly evident in voice assistants, where such performance disparities not only degrade the user experience for low-resource language users, such as those in developing regions, but also risk exacerbating the global digital divide \cite{lima2019assistants}. Beyond language, acoustic biases related to voice timbre, age, or gender can similarly degrade performance for diverse user groups. Furthermore, socio-cultural biases may be embedded within training data, causing models to perpetuate stereotypes or fail to understand context dependent on local customs and values. Technical pathways to mitigate this spectrum of biases must be equally multifaceted. This includes exploring latent biases and enhancing fairness through techniques like contrastive learning adaptations \cite{koudounas2024contrastivebias}, and developing multilingual \& multi-dialect training methods \cite{babu2022xlsr} to learn more universal audio representations. Beyond model-level techniques, fairness evaluation should go beyond average performance metrics, incorporating practices such as subgroup analysis, worst-case reporting, cross-group calibration, and robustness testing under code-switching or accented inputs, to prevent average performance from masking critical disparities in subpopulations. Ultimately, these efforts are essential for deploying truly equitable ALMs in global domains like education, healthcare, and commerce.

\subsection{Enhancing Real-World Application Readiness}
Bridging the gap from laboratory to real-world deployment requires addressing domain-specific challenges. ALMs enable voice assistants to achieve more intelligent conversational interaction and content generation. OpenAI's Realtime API integrates ASR-LLM-TTS pipelines into a unified speech-to-speech architecture, significantly reducing latency \cite{openai_realtime_api}. In healthcare, edge-deployed ALMs expand access to medical resources via on-device privacy-preserving services \cite{shukla2025healthx}. ATR improves e-commerce voice search and retrieval, handling complex product names and code-mixing. Algolia's solution integrates third-party speech transcription services (e.g., Google Cloud) to transcribe and optimize queries through data cleaning \cite{algolia_voice_search}. In customer support, AQA systems must handle multi-turn dialogue understanding and intent recognition; Meituan's WOWService employs multi-agent architectures for autonomous task management, significantly improving user satisfaction \cite{cheng2025meituan}. Despite these advances, however, broader adoption still requires further exploration of data licensing constraints, privacy requirements, latency limitations, evaluation metric uncertainty, and robustness under noise, across languages and accents.

\subsection{Developing Reliable and Comprehensive Evaluations}

Building a reliable and comprehensive evaluation ecosystem for ALMs first requires addressing fundamental reproducibility issues and data leakage. As revealed in \cite{wijngaard2024ald_survey}, YouTube-sourced datasets suffer from link rot, leading to inconsistent data compositions; meanwhile, cross-dataset contamination is prevalent. For instance, substantial unintended audio overlaps exist between WavCaps and Clotho/AudioCaps, which can inflate metrics. These issues severely compromise result comparability. Beyond these foundational concerns, ALM evaluation requires assessing capabilities beyond traditional metrics, including efficiency, safety, bias, and deployment constraints. While LLM research benefits from mature multi-dimensional benchmarks \cite{zhao2023LLMsurvey}, ALM work remains predominantly focused on cross-task performance. Although specialized evaluations for concerns like adversarial attacks are emerging \cite{song2025ajailbreak}, such benchmarks rarely simultaneously assess other helpfulness dimensions \cite{yang2025evaluationsurvey}. Cultivating unified evaluation practices that encompass these critical dimensions is essential for guiding ALMs toward robust real-world deployment.

\section{Conclusions} \label{section: Conclusions}
ALMs leverage natural language to learn fine-grained audio representations, enabling zero-shot predictions and strong performance on fine-tuned tasks. By integrating LLMs, these models demonstrate advanced processing, understanding, and reasoning capabilities for complex audio.
This survey provides the first systematic review of ALMs, integrating their development across speech, music, and natural sound within a unified audio-centric framework. We analyze the field from multiple perspectives: background, foundational architectures, pre-training, downstream transfer, data ecosystems, and evaluations. Through illustrations, formalizations, and comparative tables, we abstract common patterns in model architectures, training objectives, and evaluation methods, while mapping interdependencies between pre-training, transfer, datasets and benchmarks.
Our review further identifies and discusses core limitations such as hallucination, security vulnerabilities, bias, and high training costs.
Finally, we propose actionable directions to advance task performance and practical applications in complex audio scenes.


%





\ifCLASSOPTIONcaptionsoff
  \newpage
\fi





\bibliographystyle{IEEEtran}
\bibliography{IEEEabrv,Bibliography}

\vfill



\end{document}